\documentclass[final,5p,times,twocolumn,authoryear]{elsarticle}

\usepackage{amssymb}

\usepackage[breaklinks=true,colorlinks=true,linkcolor=blue,urlcolor=blue,citecolor=blue]{hyperref}

\begin{document}

\begin{frontmatter}

\title{Nanostructured Superconductors}

\author[inst1]{Wolfgang Lang}
\ead{wolfgang.lang@univie.ac.at}
\affiliation[inst1]{organization={University of Vienna, Faculty of Physics},
            addressline={Boltzmanngasse 5},
            city={Vienna},
            postcode={1090},
            country={Austria}}

\begin{abstract}
The relevant length scales for superconductivity are of the order of nanometers. By confining the superconducting condensate to such dimensions, many physical properties change substantially, and novel phenomena emerge, which are absent in the pristine material. We discuss various methods of creating artificial nanostructures by top-down approaches in metallic and copper-oxide superconductors and their applications. Such nanostructures can be used to control magnetic flux quanta in superconductors, anchoring them to engineered defects to avoid dissipation, guiding their motion, or building artificial flux-quanta arrangements. Nanopatterned superconductors are essential for creating model systems for basic research and enable building almost dissipationless and ultrafast electronic devices and highly sensitive sensors.
\end{abstract}


\end{frontmatter}

\tableofcontents
\vskip 0.2cm
This author-created manuscript version is made available under the \href{https://creativecommons.org/licenses/by-nc-nd/4.0}{CC-BY-NC-ND 4.0 license}. The version of record is available at \href{https://doi.org/10.1016/B978-0-323-90800-9.00014-7}{https://doi.org/10.1016/B978-0-323-90800-9.00014-7}.  \copyright\, 2024   Elsevier.
\vskip 0.2 cm
Cite as: W. Lang, ``Nanostructured Superconductors,'' in T. Chakraborty  (Ed.), Encyclopedia of Condensed Matter Physics (Second Edition). Academic Press, Oxford, 2024, pp. 368–380.

\newpage

\section{Introduction}
\label{sec:intro}
Superconductivity originates from the long-range coherence of bosonic quantum particles that condense into a lower energy state and is thus called a macroscopic quantum phenomenon. One might wonder what could be the advantage of confining the superconducting condensate to the nanoscale? The answer lies in two essential characteristic lengths that characterize superconductivity. The Ginzburg-Landau coherence length $\xi$ determines the distance over which the density of superconducting carriers can change from its peak value to zero and vice versa. The size of $\xi$ can vary from several tens of nm in metallic superconductors down to about 1\,nm in the copper-oxide superconductors with high critical temperature $T_c$ (HTS). The other length scale is set by the London penetration depth $\lambda$, which describes the decay of an external magnetic field from the edge of a superconductor toward its interior, from which it is ultimately expelled.

The ratio between these two lengths bifurcates between two types of superconductors. While so-called type-I superconductors expel a magnetic field completely (Meissner effect), the more abundant and more widely used type-II superconductors let the magnetic field enter in quantized portions of the magnetic flux $\Phi_0 = h/(2e)$, where $\Phi_0$ is termed the flux quantum, $h$ and $e$ are Planck's constant and the elementary charge, respectively. These flux quanta in a superconductor are called fluxons or vortices; the latter name results from the fact that circular supercurrents in the surrounding material stabilize these flux quanta.

Coming back to the two characteristic lengths, it turns out that the so-called Ginzburg-Landau parameter, the ratio $\kappa = \lambda/\xi$, determines the type of a superconductor, with $\kappa \geq 1/\sqrt{2}$ leading to type-II superconductivity. Another important observation is the temperature dependence of $\xi(T)=\xi(0) (1-T/T_c)^{-1/2}$ which implies a diverging enhancement of $\xi(T)$ from its minimal value $\xi(0)$ at $T = 0$ when $T_c$ is approached. Ginzburg-Landau theory also predicts the same temperature dependence also for $\lambda(T)$ and, thus, $\kappa$ is temperature-independent near $T_c$.

In what follows, we will mainly discuss thin films of superconductors. In many situations, these can be considered three-dimensional (3D) materials, as long as the coherence length perpendicular to the surface is smaller than the film's thickness $t_z$. However, from the temperature dependence of $\xi(T)$ it is evident that close to $T_c$ a transition to a two-dimensional (2D) behavior in very thin films might occur when $\xi(T) > t_z$.

Another issue becomes important for film thicknesses $t_z \lesssim \lambda(T)$. Then $\lambda(T)$ has to be replaced by an effective penetration depth (also called Pearl length) $\Lambda(T) = 2\lambda(T)^2/t_z$. Since $\lambda(T)$ (and $\Lambda(T)$, respectively) control the range of interactions between vortices, the relevant length scales can become macroscopic in very thin films. In any case, vortices separated by distances smaller than the (effective) penetration depth will behave as collective ensembles, occasionally termed `vortex matter'. Note that in a magnetic field applied perpendicular to the surface of a thin superconductor, large demagnetization effects lead to a complete penetration of the magnetic flux and to suppression of the Meissner screening.

Vortex physics is a fascinating and complex field of research, and only an introductory glimpse can be presented here to underpin the key issues related to nanostructured superconductors. For deeper insights, see \citet{BRAN24R} in this Encyclopedia and the review by \citet{BLAT94R}.

Confining the superconducting condensate to the range of the length scales introduced above inevitably means that superconductors need to be structured to dimensions of few-$\mu$m and, primarily, to the nanoscale. The use of thin films and lateral nanostructuring allows one to substantially change many physical properties and create novel phenomena, absent in the pristine material. For example, the controlled manipulation of vortices by anchoring them to engineered defects, guiding their motion, or building artificial vortex matter plays a significant role. Engineered vortex systems are essential for creating model systems for basic research and enable to build almost dissipationless and ultrafast electronic devices based on vortex manipulation, the so-called fluxonics.

Taking advantage of Josephson effects in weakly-coupled superconductors requires barriers of only a few nm in width. The Josephson junction is the central building block for sensitive magnetic field sensors, rapid single flux quantum logic circuits, THz frequency generators, and, last but not least, superconducting quantum computing.

This chapter focuses on the nanostructuring of HTS, which offer easy operation using cryocoolers or liquid nitrogen. However, the brittle nature and complex crystallographic structures of HTS make the fabrication of nanopatterns a problematic endeavor and call for new techniques. Many of these concepts have been developed since the first edition of this Encyclopedia. For a detailed overview of nanostructured metallic superconductors, the reader is referred to the book of \citet{MOSH11M}.

\section{Nanofabrication techniques}

\subsection{Lithography and ion-beam milling}
The standard techniques for nanopatterning of superconductor structures focus on thin-film processing. One of the most commonly used methods is photo- or electron-beam lithography, which is well-known for the fabrication of semiconductor devices. After growing thin films on a suitable substrate and depositing a photoresist layer on top, the planar structures are defined by illuminating the photoresist through a mask or directly processing it with an electron-beam (e-beam) writer. While the wavelength of the light limits the former method, e-beam lithography can provide resolutions at the order of 10\,nm but is a slow sequential technique.

Unfortunately, the subsequent etching processes of both photoresist and the  superconductor result in significant degradation of resolution. Another issue is that etching artifacts, such as underetching below the photoresist layer, progressively limit resolution as the thickness of the superconducting film increases. Thus, patterns in metallic superconductors are generally reported with lateral structures on the order of 100\,nm. For example, a vibrant application is the fabrication of single photon detectors \citep{GOLT01} that commonly consist of few-nm thick NbN or NbTiN films patterned to long stripes, folded into meandering patterns to increase the active area \citep{HADF09R}. The above limitations are of particular importance for the reproducible fabrication of constriction-type Josephson junctions (Dayem bridges), for which a resolution of a few nanometers would be required.

For HTS, the situation is even more complicated because they are brittle and have a complex crystallographic structure. Lithographic techniques combined with etching steps have been successfully applied to create patterns with several hundred nanometers features sizes \citep{CAST97}. The combination of e-beam lithography followed by argon-ion milling allowed the production of patterns with a minimum line width of 25\,nm \citep{SOCH10}. Since this is still too large to fabricate Josephson junctions directly, alternative fabrication methods, such as growing the thin film over a step-edge in the substrate, must be employed \citep{TAFU05R}.

\subsection{Electromigration}
An elegant way to further narrow the cross-section of superconductor bridges fabricated by conventional lithography is based on electromigration. This usually undesirable process, combining a local temperature rise and a high electric field, leads to a displacement of atoms from their original crystal-lattice positions. Closed-loop controlled electromigration, however, can avoid adverse effects. With the gradual reduction of the cross-section of aluminum constrictions to $\lesssim 150$\,nm$^2$, a geometry-induced transition from thermally-assisted phase slips to quantum phase slips has been reported \citep{BAUM16}. Other applications of electromigration include the tuning of Nb superconducting quantum interference devices (SQUIDs) \citep{COLL21} and the controlled migration of oxygen atoms in YBa$_{2}$Cu$_{3}$O$_{7}$ (YBCO) bridges \citep{MARI20}.

\subsection{Templating strategies}
In bottom-up fabrication methods, the nanostructures are already predefined during the growth process. Since patterning the substrate material is often less challenging, several methods have been developed that take advantage of the controlled imperfection in the substrate to introduce the desired structures into the superconductor. One example is the self-organized growth of highly ordered porous alumina, a membrane-like system consisting of triangular arrays of pores. A superconducting Nb thin film deposited directly on the alumina template with a pore spacing of 50\,nm acts as a nanoscale array for vortex pinning \citep{VINC07}.

Nanowires of metallic superconductors less than 10 nm in diameter and up to $1\,\mu$m in length can be fabricated by `molecular templating' \citep{BEZR00}. After a freestanding carbon nanotube is placed over a narrow and deep slit in a Si substrate, a superconductor such as MoGe is sputtered onto the entire arrangement. The resulting layer consists of the electrodes connected by the nanowire, which are all made of the same material and thus have excellent contact resistances. Subsequently, the properties of the nanowire can be further fine-tuned in a transmission electron microscope or by applying voltage pulses.

Another technique uses vicinally cut substrates with a periodic nanoscale step structure of the clean substrate surface. YBCO films grow on such substrates in a roof-tile-like arrangement of the copper-oxide layers by self-assembly. The morphology and microstructure of such vicinal films strongly depend on the miscut angle of the SrTiO$_3$ substrate and the thickness of the on-top grown YBCO \citep{HAAG97,PEDA02}, Bi$_2$Sr$_2$CaCu$_2$O$_8$ \citep{DURR00}, and Hg$_{1-x}$Re$_x$Ba$_2$CaCu$_2$O$_{6+\delta}$ (Re: rare earth element) \citep{YUN00} films. The linear arrangement of dislocations resulting from the step structure in YBCO leads to an exceptionally high critical current density. Moreover, symmetry breaking in such vicinal films also allows experimental access to out-of-plane properties of HTS, such as  resistivity, Hall effect \citep{HEIN21}, photoconductivity \citep{MARK97}, and channeling of vortex strings along the $ab$-plane \citep{BERG97}.

Strategies based on substrate modification are very versatile. Nanostructures can be deposited on the substrate prior to the deposition of a superconductor film to modify its structural and superconducting properties. A variety of interactions can be tailored with thickness modulation of the superconductor by insulating dots, proximity effects caused by metallic dots, or magnetic interactions with ferromagnetic dot arrays that interact with the vortex lattice in the superconductor \citep{MART97a}.

\subsection{Ion-induced nanostructures}
The previously discussed patterning techniques have several disadvantages, especially when applied to HTS. First, all methods that remove material result in open side faces of the HTS, allowing the relatively mobile oxygen ions to escape from the crystallographic framework and thus lowering the oxygen content. In most cases, this leads to a degradation of the superconducting properties. On the other hand, the brittle nature of HTS limits the stability of the remaining material structures, making it challenging to achieve sub-$\mu$m resolution.

Irradiation of HTS with electrons, protons, and light or heavy ions provides an alternative route to modify the properties of HTS while leaving the surface of the material nearly intact. Before going into further details, a brief account of the limiting parameters is in order. A first consideration concerns the penetration range of irradiation. Only neutrons and high-energy ions can penetrate deeply enough into bulk materials \citep{WEBE03R}. Irradiation with swift heavy ions produces columnar tracks of a few nm diameters in the HTS. The crystallographic structure is destroyed, and a non-superconducting amorphous channel remains \citep{CIVA97R}. Such a columnar defect with a diameter similar to the in-plane coherence length is ideal for blocking the motion of vortices. This enhanced pinning of vortices leads to a higher critical current. These disordered arrangements of line-shaped pinning centers are not only very important for applications but have also triggered the theoretical and experimental study of novel phases of vortex matter, such as the vortex \citep{FISH91} and Bose \citep{NELS93} glasses.

\begin{figure*}[t]
\includegraphics[width=0.7\textwidth]{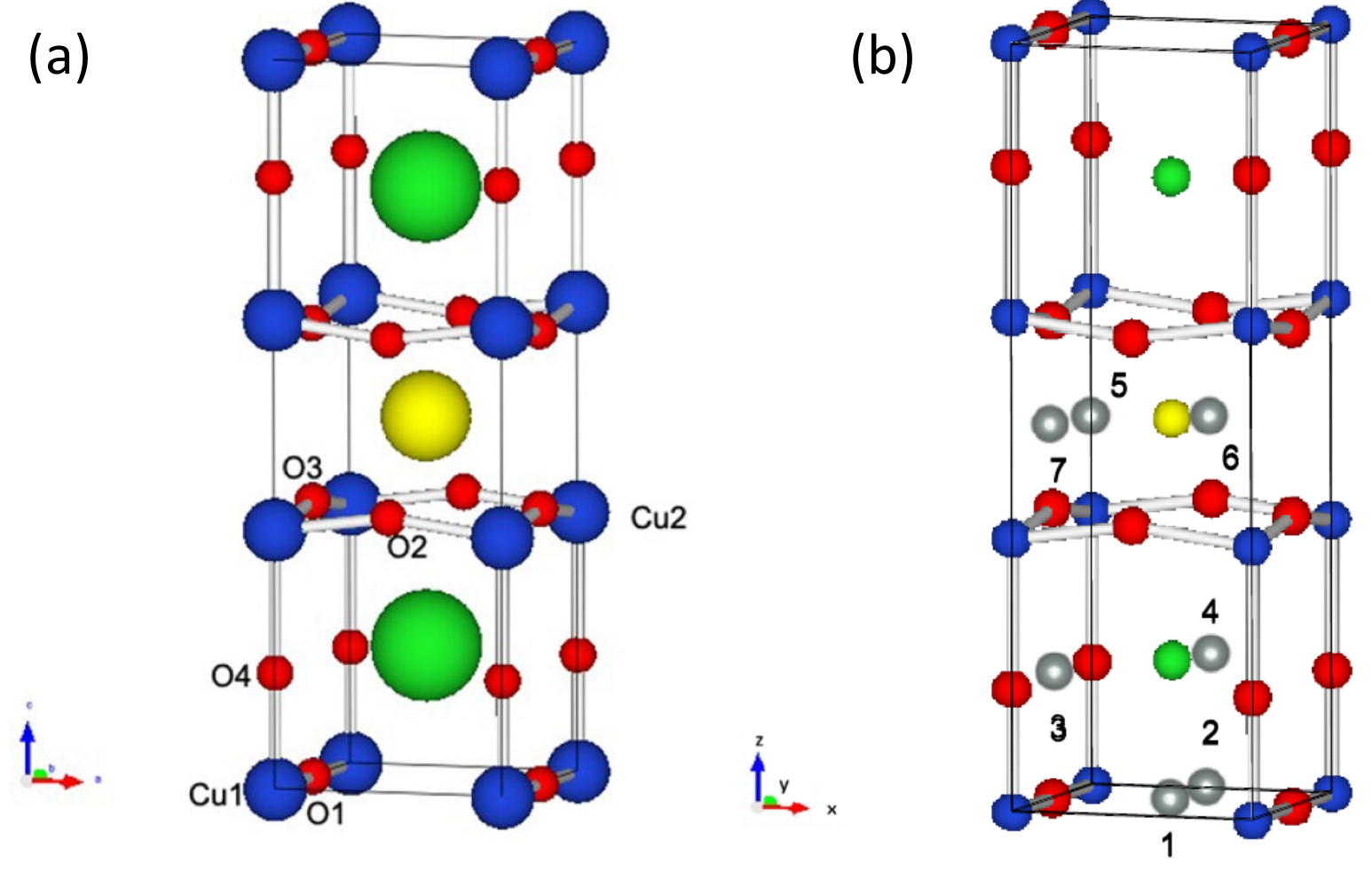}
\begin{centering}
\caption{(a) The crystal structure of the copper-oxide superconductor YBa$_{2}$Cu$_{3}$O$_{7}$. Red, blue, green, and yellow spheres correspond to oxygen, copper, barium, and yttrium. (b) The gray spheres illustrate possible interstitial sites. Figures reproduced from Gray RL, Rushton MJD, Murphy ST (2022) Molecular dynamics simulations of radiation damage in YBa$_{2}$Cu$_{3}$O$_{7}$. \emph{Superconductor Science and Technology} {\bf 35}: 035010. doi:10.1088/1361-6668/ac47dc under the \href{https://creativecommons.org/licenses/by/4.0/}{Creative Commons Attribution 4.0} licence.}
\end{centering}
\label{fig:ybco}
\end{figure*}

A more subtle method of tailoring superconducting properties is to create point defects while leaving the crystallographic framework intact. They can be produced by irradiation with electrons, protons, or light ions. Point defects are imperfections at the atomic level that occur when the energy transfer from the incident particle to the crystal is on the order of the energy required to form vacancies. At energies up to a few MeV, the incident particle collides with and displaces a nucleus, eventually creating a collisional cascade that propagates through the material. The tradeoff in the choice of parameters stems from the fact that lighter particles, such as electrons, require extremely high fluence to achieve any appreciable change of the superconducting properties. In contrast, heavier particles have limited penetration depths and more significant scattering of the collisional cascades \citep{LANG10R}.

A suitable candidate for the controlled fabrication of point defects in thin films of the most commonly used HTS, YBCO, are He$^+$ ions of moderate energy. The superconducting properties are tailored by displacing mainly oxygen atoms, which are more loosely bound than the other atomic species, with binding energies of $1 \dots 2$\,eV for the chain atoms O1 and about 8\,eV for the O2/O3 atoms in the CuO$_2$ planes, as shown in Fig.~1(a). Their displacement creates Frenkel defects at interstitial positions \citep{GRAY22} outlined in Fig.~1(b), leading to a controllable decrease \citep{SEFR01} or complete suppression of $T_c$ \citep{LANG10R}, depending on the ion fluence. Since the charge transfer from the displaced atoms is still operational, Frenkel defects do not lead to a significant change in the carrier density.

Another process is responsible for reducing $T_c$: In classical superconductors with $s$-wave symmetry of the superconducting gap, the introduction of point defects has little effect on $T_c$ and is even used technically to improve the critical current. In contrast, the anisotropic $d$-wave nature of the gap in HTS makes it susceptible to tiny defects that reduce not only the normal-state conductivity and carrier mobility but also $T_c$ \citep{LANG10R}.

\begin{figure*}[t]
\centering
\includegraphics*[width=0.9\textwidth]{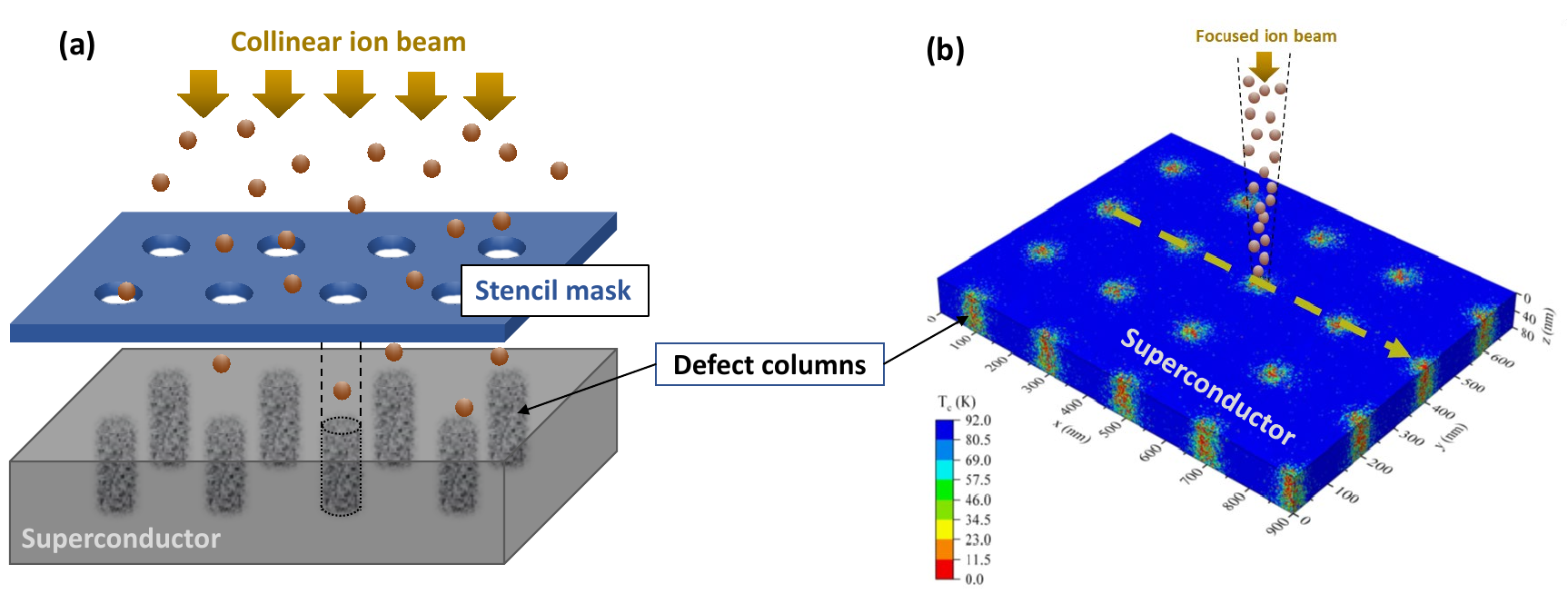}
\caption{Two different methods for patterning a YBCO film by He$^+$ ion irradiation: (a) Ion beam patterning through a stencil mask produces a large number of columnar defects in a single step. The dark areas indicate the defect-rich, non-superconducting nanocylinders. (b) Irradiation with a slightly defocused beam from a helium-ion microscope produces tailored columnar defect patterns by scanning the beam across the sample surface. The colors of the columnar defects indicate the local critical temperature, as determined by simulations. Adapted from Aichner B, Mletschnig KL, M\"uller
B, Karrer M, Dosmailov M, Pedarnig JD, Kleiner R, Koelle D, Lang W (2020) Angular magnetic-field dependence of vortex matching in pinning lattices fabricated by focused or masked helium ion beam irradiation of superconducting YBa$_{2}$Cu$_{3}$O$_{7-\delta}$ thin films. \emph{Low Temperature Physics} 46: 331–337. doi: 10.1063/10.0000863. \emph{Fiz. Nizk. Temp.} 46: 402–409.}
\label{fig:irrad}
\end{figure*}

However, a uniform statistical distribution of point defects is not very helpful. By focusing the ion beam before it hits the surface of the HTS film, one can create nearly cylindrical domains populated with point defects. These columnar defects (CDs) form a landscape where superconductivity is locally suppressed. Lateral modulation of ion fluence can be achieved mainly by two different methods outlined in Fig.~\ref{fig:irrad}.

\subsubsection{Masked ion irradiation of thin films}
An extensive array of multiple ion beams can be created by masking a wide-field collinear ion beam, which is commonly available with ion implanters, as schematically shown in Fig.~\ref{fig:irrad}(a). An HTS film, thinner than the ion's penetration depth, is fabricated on a suitable substrate. The mask protects selected areas of the HTS film from irradiation and exposes the other sample parts to irradiation. While the former remain superconducting, the $T_c$ of the latter is reduced or suppressed depending on the applied ion energy and fluence. Notably, the collision cascades widen due to the straggling of ion trajectories within the HTS so that the mask patterns become blurred with increasing depth. As a result, the lateral resolution is typically limited to about 10\,nm as indicated by simulations with 75\,keV He$^+$ irradiation \citep{HAAG14}.

Different techniques were used to create the mask. Either a photoresist layer was deposited on the HTS and processed with standard UV \citep{KAHL98}, e-beam, \citep{SWIE12}, or focused ion beam \citep{KATZ00} lithography, then etched and used as the mask, or a metal layer was deposited directly on the YBCO film and then patterned by ion beam milling \citep{KANG02a}. Alternatively, a Si stencil mask is fabricated and mounted at a small distance from the superconductor film \citep{LANG06a}. This method allows the mask to be reused, does not require the multiple processing steps associated with photoresist, and avoids potential surface degradation.

The main advantage of masking techniques is the parallel processing of many structures. Shortcomings are the resolution limitations resulting from the preparation of the mask, and in the case of freestanding masks, there are geometrical limitations, e.g., disconnected blocking elements are not possible.

\subsubsection{Focused ion beam modification of thin films}
\begin{figure*}[t]
\centering
\includegraphics*[width=0.6\textwidth]{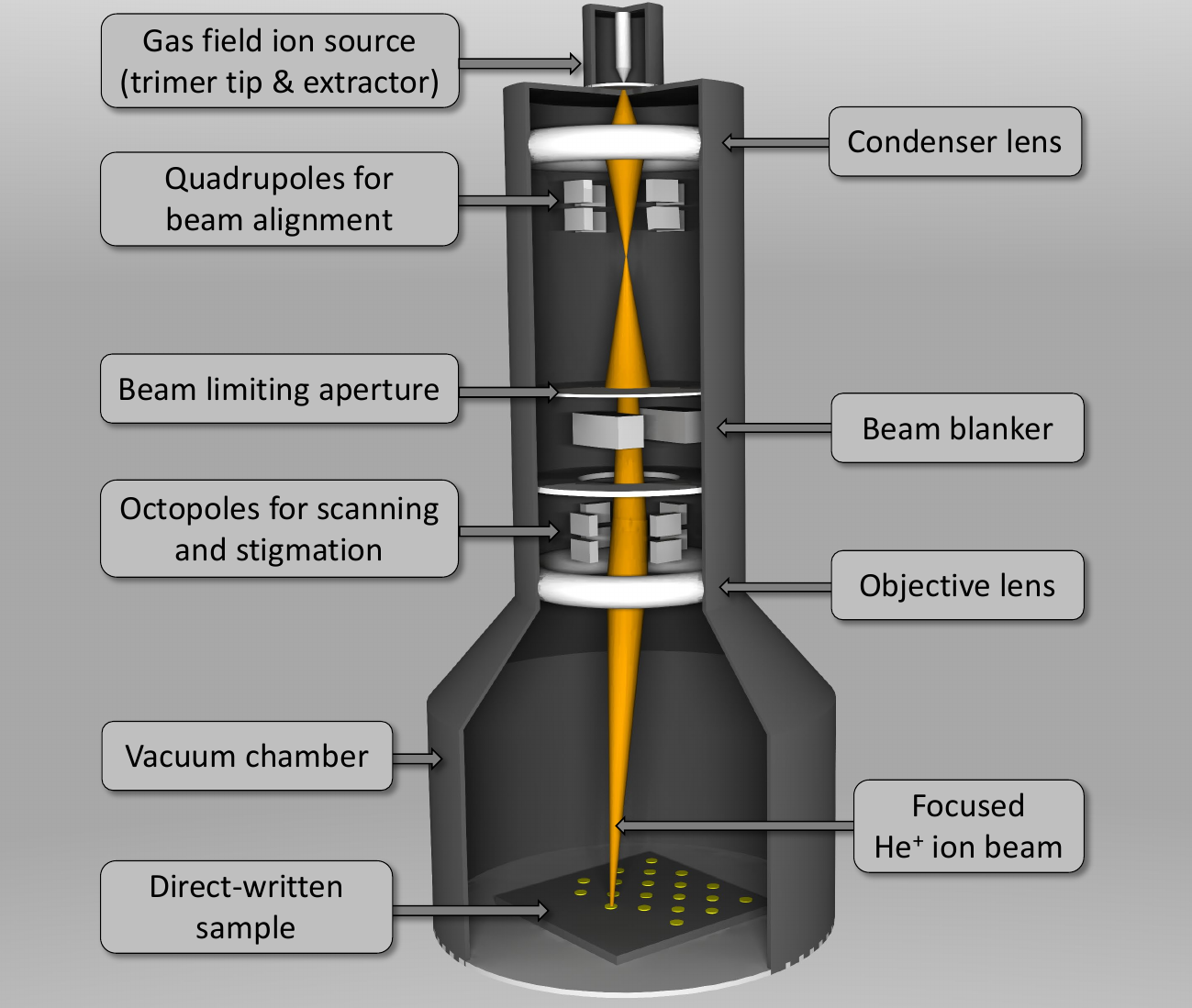}
\caption{Schematic drawing of the helium ion microscope used for direct-writing patterns of locally suppressed superconductivity. (Courtesy of Bernd Aichner, University of Vienna)}
\label{fig:him}
\end{figure*}

In contrast to conventional focused ion beam (FIB) machines, which use Ga to ablate the material, devices for focused ion irradiation with light ions have only recently become available. The helium ion microscope (HIM) \citep{WARD06} combines scanning focused ion beam sources for He$^+$ and Ne$^+$ ions and imaging via secondary electron detection. The HIM consists of a gas-field ion source that emits ions from a tip containing only three atoms (the trimer), electrostatic ion optics to focus and trim the beam, and a deflection system that moves the beam across the sample stage with optional blanking (Fig.~\ref{fig:him}). Because of the high source brightness of the trimer and the short de Broglie wavelength of He$^+$ ions, an image resolution better than 0.5\,nm and an unprecedented depth of focus can be achieved \citep{HLAW16M}.

Using a He or Ne beam instead of the conventional Ga--FIB technique, contamination of the target material by Ga ions is avoided and achieves higher lateral resolution. For example, nanopores as small as 1.3\,nm in diameter have been fabricated in 1\,nm-thick carbon nanomembranes \citep{EMMR16}. As initial applications in HTS, thin barriers of insulating material, written across prepatterned YBCO microbridges with the focused ion beam, form Josephson junctions \citep{CYBA15}, and ultradense arrays of CDs build a complex pinning landscape for vortices \citep{AICH19}. At the time of writing, hexagonal arrays of CDs with spacings as small as 20\,nm had been fabricated in YBCO thin films \citep{KARR22P}.

The use of He--FIB relies on the weak bonding of oxygen in HTS and cannot be employed in the same way for other superconductors. However, a focused Ne beam offers a compromise between Ga--FIB and He--FIB for direct milling of metallic superconductors. This has been demonstrated, for example, for patterning constrictions in NbN films to form nanowires \citep{BURN17}.

Other techniques for growing superconductor nanostructures using focused electron and ion beams are discussed in this Encyclopedia by \citet{CORD24}, and the properties of superconducting microtubes and nanohelices are examined by \citet{FOMI20M}.

\section{Properties of nanopatterned superconductors}
Nanostructuring of superconductors enables selective tailoring of the superconducting condensate on length scales smaller than the London penetration depth, mainly by creating lateral structures. Such patterns allow controlled interaction of vortices, their manipulation, as well as tunneling effects between two superconducting condensates weakly coupled by a small non-superconducting interlayer. Some of these applications are described below.

\subsection{Vortex pinning arrays}
Lateral nanostructuring of superconducting films with regular arrays of CDs allows the creation of artificial pinning landscapes that lead to commensurability effects with the flux line lattice. This occurs at the so-called matching fields $B_k = k \Phi _0/A$, where $k$ is an integer number of pinning sites (or a rational number of vortices) in the unit cell of the pinning array of area $A$. For square lattices $A = a^2$ and for hexagonal patterns $A = \sqrt{3} a^2/2$ with $a$ the nearest neighbor spacing of the CDs.

\begin{figure*}[t]
\centering
\includegraphics*[width=\textwidth]{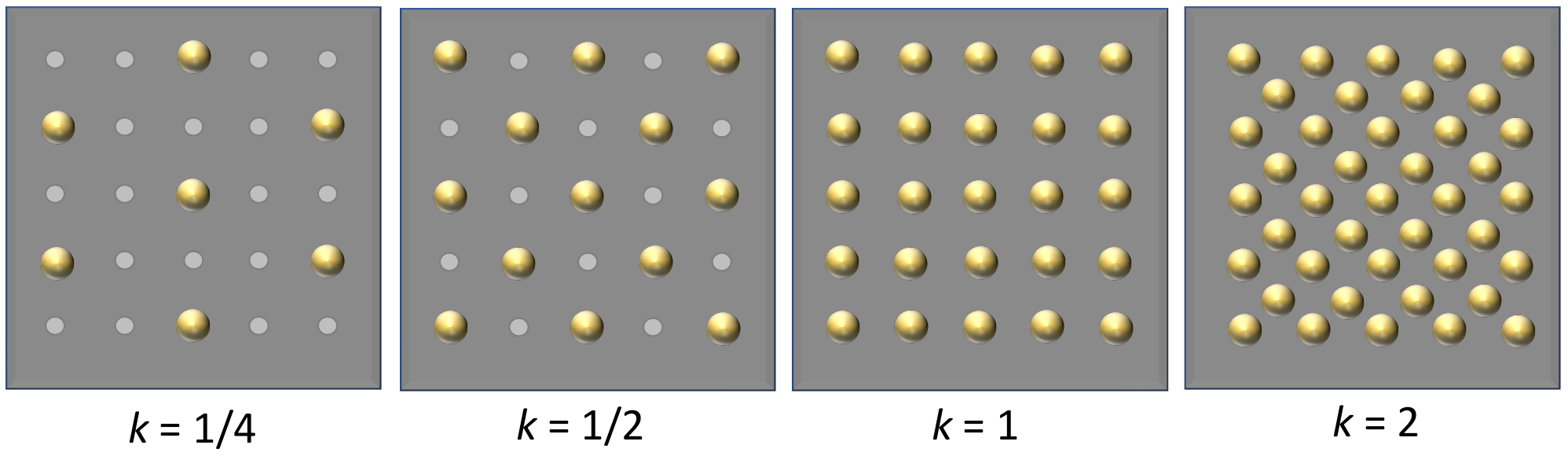}
\caption{Schematic representation of different commensurability patterns observed by Lorentz microscopy. Light gray circles represent holes in the superconducting film and yellow bullets represent the magnetic flux quanta as observed by a deflection of the electron beam. After \citet{HARA96b}.}
\label{fig:match}
\end{figure*}

The number of flux quanta that can be trapped inside a cylindrical hole or in a non-superconducting pinning site depends on its diameter $r_p$ and the coherence length $\xi(T)$ of the superconductor. For anisotropic layered materials such as HTS, the in-plane coherence length $\xi_{ab}(T)$ is relevant. Once all pinning sites are filled by one flux quantum, the formation of multiquanta vortices becomes energetically favorable when $r_p$ exceeds a certain critical radius \citep{BUZD93}. The saturation number of trapped flux quanta can be estimated by $n_S \simeq r_p/2\xi(T)$ \citep{BUZD96b}.

A special situation arises for blind holes in a superconductor, i.e., holes that do not completely penetrate the material. The remaining superconducting bottom layer carries the screening currents of individual vortices trapped in the blind hole. Although several flux quanta are trapped in the blind hole, individual vortices can still be made visible \citep{BEZR96}.

Multiquanta vortices are a phenomenon that occurs only in nanostructured superconductors and does not exist in pure homogeneous superconductors. Conversely, when small pinning sites cannot accommodate the total flux, the excess flux is forced to enter the material via interstitial vortices. The various commensurability effects in a pinning landscape for different magnetic fields applied orthogonally to the film surface are illustrated in Fig.~\ref{fig:match}, which is based on experimental results by Lorentz microscopy of vortices in a Pb film patterned with a square array of tiny blind holes. For filling factors $k < 1$, the flux quanta are arranged in a superlattice with respect to the pinning array; for $k = 1$, each hole is filled with precisely one flux quantum, and for $k = 2$, the excess flux is taken up by interstitial vortices.

\begin{figure*}[t]
\centering
\includegraphics*[width=0.8\textwidth]{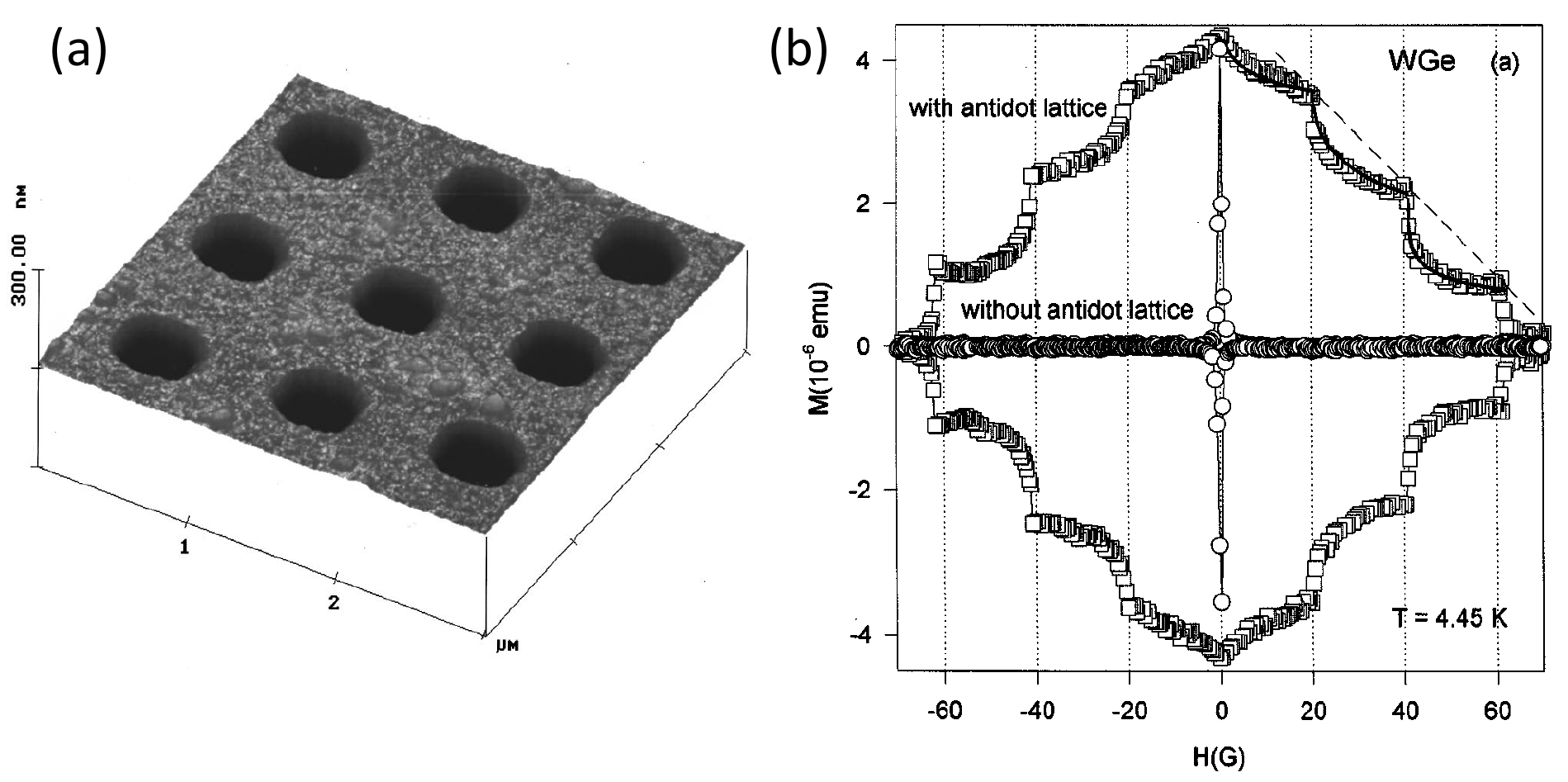}
\caption{(a) Atomic force microscopy image of a 60\,nm-thick WGe film with a square lattice of holes (antidots) with 340\,nm diameter and $1\,\mu$m spacing. (b) Magnetization loop $M(H)$ at $T = 4.45$\,K of the WGe films with and without the antidot grid. The dotted vertical lines indicate the matching fields $B_k$, $k = 1,2,3$. Figures reprinted with permission from Moshchalkov VV, Baert M, Metlushko VV, Rosseel E, Bael MJV, Temst K, Jonckheere R and Bruynseraede Y (1996) \emph{Physical Review B} 54: 7385. Copyright 1996 by the American Physical Society.}
\label{fig:magnetization}
\end{figure*}

The commensurability effects manifest themselves in various physical parameters, such as steps in the magnetization loops and peaks in the critical current as indicators of enhanced pinning forces. Minima in the resistance versus magnetic field curves point to commensurability effects of moving vortex ensembles.

Fig.~\ref{fig:magnetization} shows an example of the dramatic change of the magnetization loop $M(H)$ after perforating a 60\,nm-thick WGe film with a square array of holes 340\,nm in diameter and $1\,\mu$m apart  \citep{MOSH96a}. The area of the $M(H)$ loop in the perforated film is massively increased due to the holes acting as artificial pinning centers. Distinct cusps in the loop appear at the matching fields $B_k = k \times 2.07$\,mT, indicating the trapping of multiquanta vortices. The staircase-like reduction of $M(H)$ with increasing field results from the few excess vortices that appear after filling the holes with $k$ vortices. These are initially repelled by the trapped vortices and can move in the interstitial region with higher mobility. As soon as the number of the excess vortices is large enough, they fill up the holes to $k+1$ multiquanta vortices.

When the diameter of holes in a superconducting film is increased to a size close to their spacing, a gradual change from a pinning lattice to a multiply connected network of superconducting wires takes place. Such wire networks exhibit a modulation of the critical temperature with the magnetic flux, the Little-Parks effect \citep{PARK64}. A further enlargement of the holes leads to the formation of disconnected superconducting islands with intriguing properties \citep{POCC15}.

\begin{figure*}[t!]
\centering
\includegraphics*[width=0.9\textwidth]{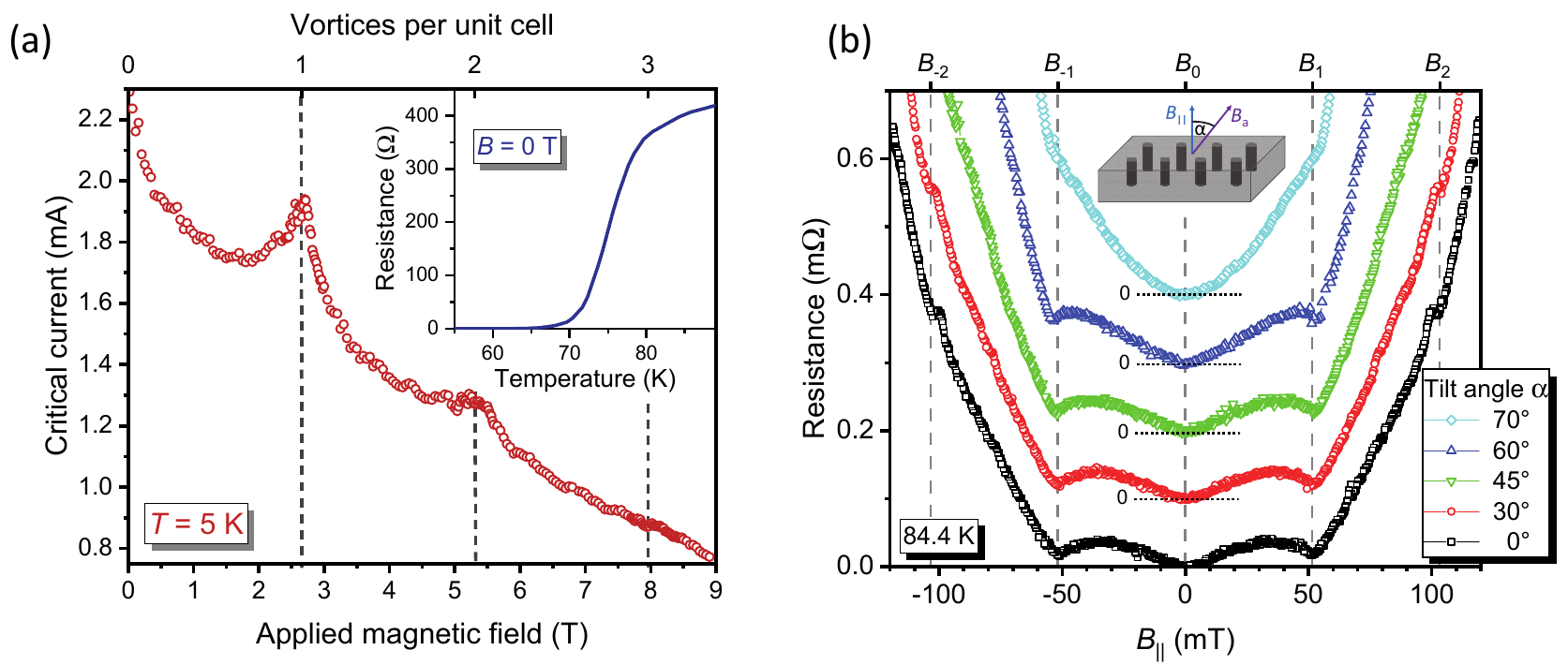}
\caption{Vortex commensurability effects in YBCO thin films with an array of columnar defects, created by focused He$^+$ ion beam irradiation. (a) Critical current at 5\,K as a function of applied magnetic field of a thin YBCO film with a hexagonal pattern of columnar defects spaced 30\,nm apart. At the first matching field $B_1 = 2.65$\,T, a peak of the critical current is caused by strong commensurability effects. Another peak occurs when two vortices are trapped in the pins. The inset shows the superconducting transition in zero magnetic field. Adapted from \citet{BACK22}. (b) Angular dependence of vortex matching effects in the resistance of a YBCO thin film with a square lattice of 200\,nm spaced columnar defects: The resistance is plotted as a function of the applied field component along the normal of the film surface $B_\parallel = B_a \cos \alpha$ for different values of $\alpha$. For $\alpha > 0^\circ$ the curves are shifted by multiples of 0.1\,m$\Omega$ for better visibility. The inset shows a sketch of the experimental situation. Reprinted with permission from \citet{AICH20}.}
\label{fig:commens}
\end{figure*}

In HTS, commensurability effects can be preferably detected by electric transport measurements, e.g., in YBCO thin films patterned with a square lattice of holes \citep{CAST97}. However, the situation is more complicated because YBCO thin films have strong inherent pinning by crystallographic microtwinning, screw dislocations, and other intrinsic defects. The competition between pinning on these immanent defects and trapping vortices on engineered pinning sites requires that the latter be relatively dense, with a spacing of $\lesssim 300$\,nm. Such a resolution is hard to achieve by lithographic methods but is within reach of masked or focused ion beam modification, which can raise the magnitude of the matching fields into the range of several tesla.

A demonstration of vortex commensurability effects at high magnetic fields is shown in Fig~\ref{fig:commens}(a) in a YBCO film with a dense hexagonal pinning lattice with $a = 30$\,nm spacing. Since $a$ is much smaller than the Pearl length $\Lambda(0)$, peaks in the critical current caused by enhanced pinning of vortices can be observed at temperatures far below the superconducting transition. Matching effects of mobile vortices at higher temperatures can be traced as dips in the resistivity, as shown in Fig.~\ref{fig:commens}(b). Moreover, the preferential trapping of vortices within the CDs can be confirmed by tilting the applied magnetic field away from the axes of the CDs by an angle $\alpha$ as illustrated in the inset of Fig.~\ref{fig:commens}(b). Then, the position of the matching dips scales with the magnetic field component parallel to the axes of the CDs. These effects prevail up to high tilt angles of $\alpha \leq 70^\circ$ \citep{AICH20}.

\begin{figure*}[t]
\centering
\includegraphics*[width=\textwidth]{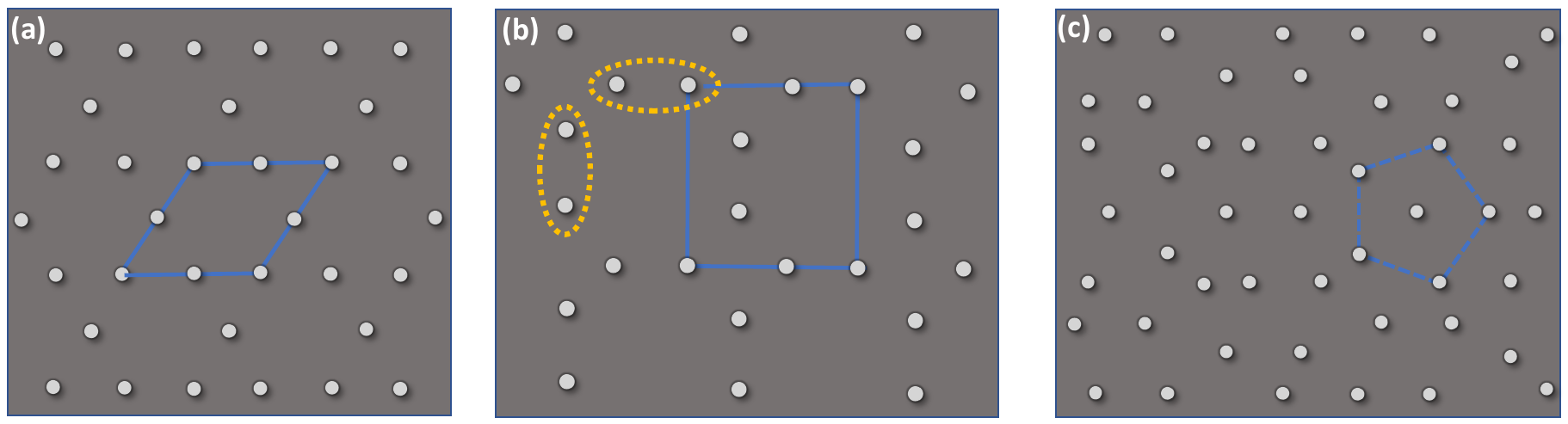}
\caption{Examples of complex pinning landscapes. Bright circles indicate pinning centers (holes or columnar defects) and solid blue lines the unit cell of a periodic lattice. (a) The Kagom\'e lattice is related to the hexagonal lattice. The latter can be recovered if one pin is added in the center of each void. (b) Vortex-ice lattice: four pairs of pins (indicated by orange ellipses) meet at each vertex of a square lattice. (c) The aperiodic Penrose tiling. The dashed blue lines indicate the five-fold symmetry.}
\label{fig:tilings}
\end{figure*}

\subsection{Complex pinning landscapes}
When moving from pinning lattices with hexagonal or square arrangements to more complex periodic or aperiodic tilings, numerous unusual phenomena occur. Some examples are shown in Fig.~\ref{fig:tilings}. Such patterns have been studied theoretically \citep{LAGU01,MISK05,REIC07a,MISK12} and experimentally in metallic superconductors with holes \citep{KEMM06,MISK10,BOTH14} and magnetic dots \citep{VILL06,SILH06,KRAM09}, like Penrose \citep{KEMM06,MISK10}, honeycomb \citep{WELP02,LATI12} and Kagom\'e \citep{CUPP11} lattices and artificial vortex ice arrangements in geometrically frustrated pinning lattices \citep{LIBA09,LATI13,XUE18}.

In HTS, such studies are more challenging and could be hampered by strong intrinsic pinning. However, complex pinning structures lead to competition between the pinning forces at the CDs and the elastic energy of the vortex lattice, attempting to restore the natural hexagonal vortex configuration of a clean superconductor. For example, pinning landscapes that force the magnetic flux quanta in an ice-like flux arrangement due to a geometrically frustrated energy landscape can transition to a periodic flux distribution at higher temperatures, thawing the vortex ice \citep{TRAS14}.

Another example of a complex pinning landscape, the quasi-Kagom\'e tiling, shows an unconventional commensurability effect. At elevated temperatures, all pins are occupied by vortices, and one interstitial vortex is magnetically caged in each void of the lattice. The balance between the pinning forces exerted by the CDs and the vortex caging potential can be tuned with the temperature \citep{AICH19}. Controlled manipulation of such magnetically confined vortices can pave the way toward fluxonic applications of HTS.

\subsection{Vortex ratchets and flux diodes}
By introducing spatial asymmetry into the pinning arrays, many exciting effects emerge. In general, the idea of a `Brownian motor' refers to Brownian motion in combination with an unbiased external input signal that can induce submicron directional motion of particles \citep{HANG09R}. A well-known example is the directional propagation of vortices in an appropriately structured superconductor, usually referred to as a `vortex ratchet' or `fluxon pump' \citep{WAMB99}. It provides a flexible and well-controlled model system for studying  stochastic transport processes and can operate up to THz frequencies \citep{HAST03}.

Vortex ratchets can be realized by various concepts, such as 2D asymmetric channel walls. They can be further extended to design `fluxon optics' devices, concave/convex fluxon lenses that disperse/concentrate fluxons in nanodevices \citep{WAMB99}. Also, asymmetric potential barriers, e.g., in the form of square arrays of triangular pinning centers \citep{VILL03}, double-well traps, \citep{VAND05} or an asymmetric arrangement of symmetric traps, \citep{WORD04,LARA10} lead to vortex rectification effects.

Interestingly, ratchet effects are proposed in binary particle mixtures without needing for an asymmetric substrate \citep{SAVE02}. For example, in layered superconductors, such as HTS, an inclination of the magnetic field from the $c$ axis leads to a mixture of pancake vortices and Josephson strings. This hybrid vortex system exhibits ratchet effects with time asymmetric drives \citep{COLE06}.

Finally, understanding ratchet effects in different systems is an exciting topic. Studying these effects with fluxons provides a more direct and controllable experimental approach than most other systems.  Ultimately, these investigations may pave the way to cellular automata as an alternative concept \citep{HAST03,MILO07} for performing clocked logic operations on discrete particles. Moreover, vortex ratchets are proposed \citep{LEE99} as an effective method for evacuating magnetic flux from superconducting devices where inadvertently trapped flux might be detrimental for operation, such as in SQUID magnetic sensors.

A related concept is the lossless superconducting diode. This is an electronic device that has zero resistance only for one direction of applied current and is a desirable device for building electronic circuits with ultra-low power consumption. It can be realized as a superconducting film patterned with a conformal array of nanoscale holes that breaks spatial inversion symmetry \citep{LYU21}. A conformal array is a structure resulting from the transformation of a uniform hexagonal pinning array that retains the sixfold order of the original lattice but exhibits a gradient in site density \citep{REIC15}.

\subsection{Guided vortex motion}
Controlled vortex motion along a predefined path can be achieved by patterning narrow channels or parallel rows of defects into a superconductor. This so-called `guided vortex motion' results from an easy track in which vortex pinning is reduced \citep{DOBR16}, a row of holes \citep{SILH10R}, or, in HTS, the suppression of superconductivity by heavy-ion irradiation \citep{LAVI10}. Guided vortex motion can be detected by the deflection of vortices from their trajectory imposed by the Lorentz force.  This leads to a pronounced transverse voltage \citep{WORD04} that can be distinguished from the conventional Hall effect by its even symmetry with respect to the reversal of the magnetic field. Guided vortex motion also plays an important role in experiments where vortices are accelerated to high velocities, as discussed by \citet{DOBR24R}.

\subsection{Josephson junctions}
A ubiquitous need for nanostructuring of superconductors arises from the fabrication of Josephson junctions (JJs), where two superconducting systems are separated by an insulating or metallic barrier of a few nm thickness or by another weak coupling link. It is crucial that the Josephson weak links are stable and can be reproducibly fabricated. While for metallic superconductors the industrial fabrication of circuits based on JJs has been established for many years \citep{LIKH12R}, the fabrication of JJs in HTS is much more challenging. It is still in the phase of cumulative progress.

JJs consisting exclusively of HTS materials have been produced by introducing a crystallographic fault (break junctions), by using a grain boundary between different crystallites, by growing thin HTS films over a substrate step, by narrow constrictions (Dayem bridges), or by multilayers forcing the current along the $c$-axis \citep{KOEL99R}. Here we restrict ourselves to discussing JJs in HTS produced by masked or focused ion irradiation. A general account of this subject can be found in the chapter by \citet{TAFU24R}.

Several attempts have been made to fabricate JJs by irradiation techniques. Initially, direct writing of narrow lines with an electron beam across a pre-patterned YBCO bridge in a scanning electron microscope resulted in somewhat unstable JJs and required high electron doses. Later, creating a weak link in YBCO bridges by implanting oxygen ions through a lithographically defined mask led to superconducting-normal-superconducting (SNS) JJs with resistively shunted junction (RSJ) properties \citep{TINC96,KAHL98,BERG05}. However, the minimum width of the mask structures of about 20\,nm, and the inevitable straggle of ion collision cascades in the YBCO film set resolution limits and prevent fabrication of superconductor-insulator-superconductor (SIS) junctions, which would be the ultimate goal.

A similar technique, but using 200\,keV Ne$^+$ ions, allows complete penetration of the ions through the YBCO film, avoiding implantation and limiting the intended damage to the creation of point defects \citep{KATZ98}. This technique can be scaled up to integrating of many JJs in a 2D series-parallel array; 15,820 ($28 \times 565$) JJs have been demonstrated \citep{CYBA09}.

\begin{figure*}[t!]
\centering
\includegraphics*[width=0.85\textwidth]{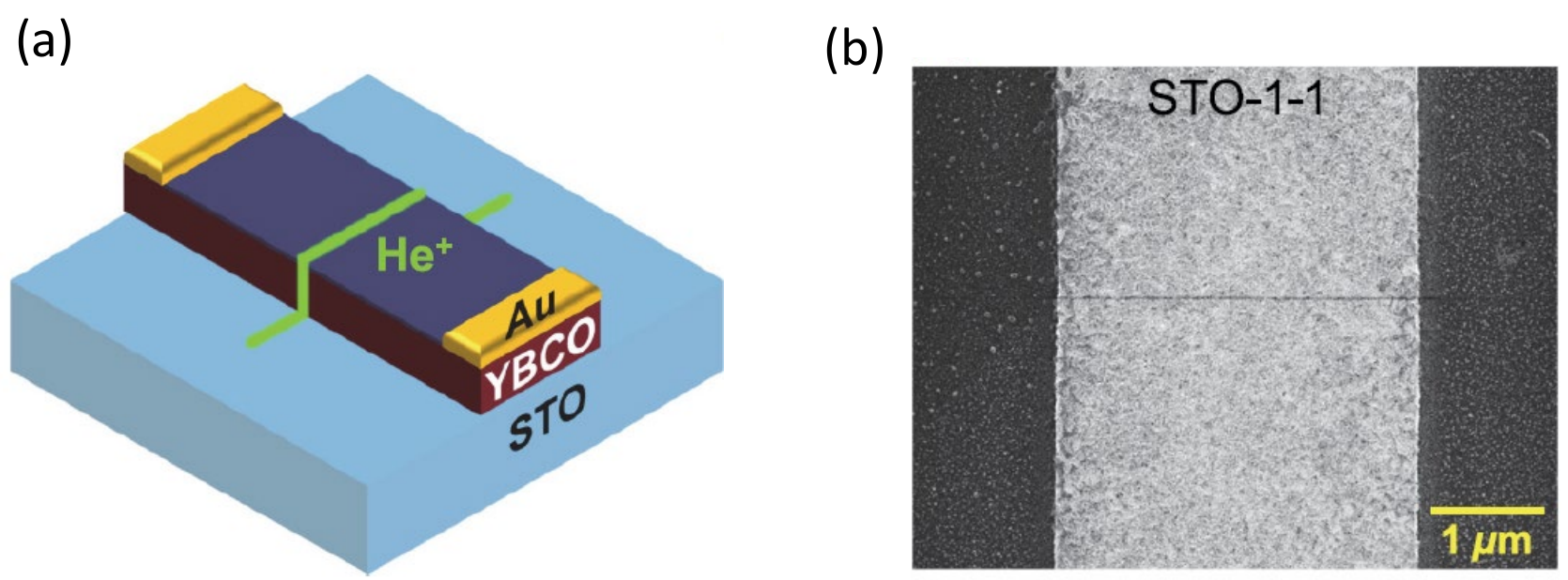}
\caption{(a) Schematic illustration of a Josephson junction fabricated by focused He$^+$ ion irradiation in a helium ion microscope. (b) Scanning electron image of the junction (visible as a thin dark line) patterned with a fluence of 600\,ions/nm. Figures reprinted with permission from M\"uller B, Karrer M, Limberger F, Becker M, Schr\"oppel B, Burkhardt CJ, Kleiner R, Goldobin E and Koelle D (2019) \emph{Physical Review Applied} 11: 044082. Copyright 2019 by the American Physical Society.}
\label{fig:jj}
\end{figure*}

Focused He$^+$ ion irradiation in a HIM took the fabrication of JJs one step further. Tunnel junctions can be directly written with the focused He-ion beam into YBCO films, as illustrated in Fig.~\ref{fig:jj}. The properties of the barrier are controlled by varying the irradiation dose. With this technique, SIS junction can also be realized \citep{CYBA14}. Scanning transmission electron analysis shows that the amorphous tracks created by 1500 He$^+$/nm have a lateral extension of 4\,nm, while no destruction of the crystallographic structure is observed at lower fluence. Nevertheless, the devices produced with the lower doses show an explicit JJ behavior \citep{MULL19}.

The He--FIB technique can, in principle, be applied to other HTS, as has been demonstrated for La$_{1.84}$Sr$_{0.16}$CuO$_4$ \citep{GOZA17} and other superconducting materials such as MgB$_2$ \citep{KASA18}. The versatility of directly written structures and JJs provides a tremendous advantage in fabricating superconducting quantum interference devices \citep{MULL19}, JJ arrays \citep{LEFE19}, and even more complex devices on the same substrate for various applications.

\section{Conclusions and outlook}
In this chapter, we have outlined various methods for fabricating nanostructured superconductors. While established lithographic and ion-milling techniques enable the patterning of metallic superconductors, novel techniques are required for nanoscale structures in copper-oxide superconductors. For example, focused He$^+$-ion beam irradiation creates columnar channels of point defects and can be used to create vortex pinning landscapes and weak links for Josephson junctions. Superconducting nanostructures are already indispensable for many applications and will bring further significant advances to applications of superconducting electronics, from fluxonics to ultra-high sensitivity magnetometers and many others. Beyond that, the dawning age of superconducting quantum computing depends heavily on precise and reproducible nanostructured circuits on a large scale.

\section*{Acknowledgments}
This work was supported by the Austrian Science Fund (FWF), grant I4865-N, and is based upon work from COST Actions CA16218 (NANOCOHYBRI), CA19108 (Hi-SCALE), and CA19140 (FIT4NANO), supported by COST (European Cooperation in Science and Technology).

\addcontentsline{toc}{section}{References}
\bibliographystyle{elsarticle-harv}

\begin{thebibliography}{98}
\expandafter\ifx\csname natexlab\endcsname\relax\def\natexlab#1{#1}\fi
\providecommand{\url}[1]{\texttt{#1}}
\providecommand{\href}[2]{#2}
\providecommand{\path}[1]{#1}
\providecommand{\DOIprefix}{doi:}
\providecommand{\ArXivprefix}{arXiv:}
\providecommand{\URLprefix}{URL: }
\providecommand{\Pubmedprefix}{pmid:}
\providecommand{\doi}[1]{\href{http://dx.doi.org/#1}{\path{#1}}}
\providecommand{\Pubmed}[1]{\href{pmid:#1}{\path{#1}}}
\providecommand{\bibinfo}[2]{#2}
\ifx\xfnm\relax \def\xfnm[#1]{\unskip,\space#1}\fi
\bibitem[{Aichner et~al.(2020)Aichner, Mletschnig, M\"uller, Karrer, Dosmailov,
  Pedarnig, Kleiner, Koelle and Lang}]{AICH20}
\bibinfo{author}{Aichner, B.}, \bibinfo{author}{Mletschnig, K.L.},
  \bibinfo{author}{M\"uller, B.}, \bibinfo{author}{Karrer, M.},
  \bibinfo{author}{Dosmailov, M.}, \bibinfo{author}{Pedarnig, J.D.},
  \bibinfo{author}{Kleiner, R.}, \bibinfo{author}{Koelle, D.},
  \bibinfo{author}{Lang, W.}, \bibinfo{year}{2020}.
\newblock \bibinfo{title}{Angular magnetic-field dependence of vortex matching
  in pinning lattices fabricated by focused or masked helium ion beam
  irradiation of superconducting {YBa$_{2}$Cu$_{3}$O$_{7-\delta}$} thin films}.
\newblock \bibinfo{journal}{Low Temp. Phys.} \bibinfo{volume}{46},
  \bibinfo{pages}{331--337}.
\newblock \DOIprefix\doi{10.1063/10.0000863}. \bibinfo{note}{{F}iz. Nizk. Temp.
  {\bf 46}, 402-–409}.
\bibitem[{Aichner et~al.(2019)Aichner, M\"uller, Karrer, Misko, Limberger,
  Mletschnig, Dosmailov, Pedarnig, Nori, Kleiner, Koelle and Lang}]{AICH19}
\bibinfo{author}{Aichner, B.}, \bibinfo{author}{M\"uller, B.},
  \bibinfo{author}{Karrer, M.}, \bibinfo{author}{Misko, V.R.},
  \bibinfo{author}{Limberger, F.}, \bibinfo{author}{Mletschnig, K.L.},
  \bibinfo{author}{Dosmailov, M.}, \bibinfo{author}{Pedarnig, J.D.},
  \bibinfo{author}{Nori, F.}, \bibinfo{author}{Kleiner, R.},
  \bibinfo{author}{Koelle, D.}, \bibinfo{author}{Lang, W.},
  \bibinfo{year}{2019}.
\newblock \bibinfo{title}{Ultradense tailored vortex pinning arrays in
  superconducting {YBa}$_2${Cu}$_3${O}$_{7-\delta}$ thin films created by
  focused {He} ion beam irradiation for fluxonics applications}.
\newblock \bibinfo{journal}{{ACS} Appl. Nano Mater.} \bibinfo{volume}{2},
  \bibinfo{pages}{5108--5115}.
\newblock \DOIprefix\doi{10.1021/acsanm.9b01006}.
\bibitem[{Backmeister et~al.(2022)Backmeister, Aichner, Karrer, Wurster,
  Kleiner, Goldobin, Koelle and Lang}]{BACK22}
\bibinfo{author}{Backmeister, L.}, \bibinfo{author}{Aichner, B.},
  \bibinfo{author}{Karrer, M.}, \bibinfo{author}{Wurster, K.},
  \bibinfo{author}{Kleiner, R.}, \bibinfo{author}{Goldobin, E.},
  \bibinfo{author}{Koelle, D.}, \bibinfo{author}{Lang, W.},
  \bibinfo{year}{2022}.
\newblock \bibinfo{title}{Ordered {B}ose glass of vortices in superconducting
  {YBa$_{2}$Cu$_{3}$O$_{7-\delta}$} thin films with a periodic pin lattice
  created by focused helium ion irradiation}.
\newblock \bibinfo{journal}{Nanomaterials} \bibinfo{volume}{12},
  \bibinfo{pages}{3491}.
\newblock \DOIprefix\doi{10.3390/nano12193491}.
\bibitem[{Baumans et~al.(2016)Baumans, Cerbu, Adami, Zharinov, Verellen,
  Papari, Scheerder, Zhang, Moshchalkov, Silhanek and {Van~de~Vondel}}]{BAUM16}
\bibinfo{author}{Baumans, X.D.A.}, \bibinfo{author}{Cerbu, D.},
  \bibinfo{author}{Adami, O.A.}, \bibinfo{author}{Zharinov, V.S.},
  \bibinfo{author}{Verellen, N.}, \bibinfo{author}{Papari, G.},
  \bibinfo{author}{Scheerder, J.E.}, \bibinfo{author}{Zhang, G.},
  \bibinfo{author}{Moshchalkov, V.V.}, \bibinfo{author}{Silhanek, A.V.},
  \bibinfo{author}{{Van~de~Vondel}, J.}, \bibinfo{year}{2016}.
\newblock \bibinfo{title}{Thermal and quantum depletion of superconductivity in
  narrow junctions created by controlled electromigration}.
\newblock \bibinfo{journal}{Nat. Commun.} \bibinfo{volume}{7},
  \bibinfo{pages}{10560}.
\newblock \DOIprefix\doi{10.1038/ncomms10560}.
\bibitem[{Bergeal et~al.(2005)Bergeal, Grison, Lesueur, Faini, Aprili and
  Contour}]{BERG05}
\bibinfo{author}{Bergeal, N.}, \bibinfo{author}{Grison, X.},
  \bibinfo{author}{Lesueur, J.}, \bibinfo{author}{Faini, G.},
  \bibinfo{author}{Aprili, M.}, \bibinfo{author}{Contour, J.P.},
  \bibinfo{year}{2005}.
\newblock \bibinfo{title}{High-quality planar high-{$T_c$} {J}osephson
  junctions}.
\newblock \bibinfo{journal}{Appl. Phys. Lett.} \bibinfo{volume}{87},
  \bibinfo{pages}{102502}.
\newblock \DOIprefix\doi{10.1063/1.2037206}.
\bibitem[{Berghuis et~al.(1997)Berghuis, Di~Bartolomeo, Wagner and
  Evetts}]{BERG97}
\bibinfo{author}{Berghuis, P.}, \bibinfo{author}{Di~Bartolomeo, E.},
  \bibinfo{author}{Wagner, G.A.}, \bibinfo{author}{Evetts, J.E.},
  \bibinfo{year}{1997}.
\newblock \bibinfo{title}{Intrinsic channeling of vortices along the $ab$ plane
  in vicinal {YBa$_{2}$Cu$_{3}$O$_{7-\delta}$} films}.
\newblock \bibinfo{journal}{Phys. Rev. Lett.} \bibinfo{volume}{79},
  \bibinfo{pages}{2332--2335}.
\newblock \DOIprefix\doi{10.1103/PhysRevLett.79.2332}.
\bibitem[{Bezryadin et~al.(2000)Bezryadin, Lau and Tinkham}]{BEZR00}
\bibinfo{author}{Bezryadin, A.}, \bibinfo{author}{Lau, C.N.},
  \bibinfo{author}{Tinkham, M.}, \bibinfo{year}{2000}.
\newblock \bibinfo{title}{Quantum suppression of superconductivity in ultrathin
  nanowires}.
\newblock \bibinfo{journal}{Nature} \bibinfo{volume}{404},
  \bibinfo{pages}{971--974}.
\newblock \DOIprefix\doi{10.1038/35010060}.
\bibitem[{Bezryadin et~al.(1996)Bezryadin, Ovchinnikov and Pannetier}]{BEZR96}
\bibinfo{author}{Bezryadin, A.}, \bibinfo{author}{Ovchinnikov, Y.N.},
  \bibinfo{author}{Pannetier, B.}, \bibinfo{year}{1996}.
\newblock \bibinfo{title}{Nucleation of vortices inside open and blind
  microholes}.
\newblock \bibinfo{journal}{Phys. Rev. B} \bibinfo{volume}{53},
  \bibinfo{pages}{8553--8560}.
\newblock \DOIprefix\doi{10.1103/PhysRevB.53.8553}.
\bibitem[{Blatter et~al.(1994)Blatter, Feigel'man, Geshkenbein, Larkin and
  Vinokur}]{BLAT94R}
\bibinfo{author}{Blatter, G.}, \bibinfo{author}{Feigel'man, M.V.},
  \bibinfo{author}{Geshkenbein, V.B.}, \bibinfo{author}{Larkin, A.I.},
  \bibinfo{author}{Vinokur, V.M.}, \bibinfo{year}{1994}.
\newblock \bibinfo{title}{Vortices in high-temperature superconductors}.
\newblock \bibinfo{journal}{Rev. Mod. Phys.} \bibinfo{volume}{66},
  \bibinfo{pages}{1125--1388}.
\newblock \DOIprefix\doi{10.1103/RevModPhys.66.1125}.
\bibitem[{Bothner et~al.(2014)Bothner, Seidl, Misko, Kleiner, Koelle and
  Kemmler}]{BOTH14}
\bibinfo{author}{Bothner, D.}, \bibinfo{author}{Seidl, R.},
  \bibinfo{author}{Misko, V.R.}, \bibinfo{author}{Kleiner, R.},
  \bibinfo{author}{Koelle, D.}, \bibinfo{author}{Kemmler, M.},
  \bibinfo{year}{2014}.
\newblock \bibinfo{title}{Unusual commensurability effects in quasiperiodic
  pinning arrays induced by local inhomogeneities of the pinning site density}.
\newblock \bibinfo{journal}{Supercond. Sci. Technol.} \bibinfo{volume}{27},
  \bibinfo{pages}{065002}.
\newblock \DOIprefix\doi{10.1088/0953-2048/27/6/065002}.
\bibitem[{Brandt(2024)}]{BRAN24R}
\bibinfo{author}{Brandt, E.H.}, \bibinfo{year}{2024}.
\newblock \bibinfo{title}{Superconductivity: {Ginzburg}-{Landau} theory and
  vortex lattice}, in: \bibinfo{editor}{Chakraborty, T.} (Ed.),
  \bibinfo{booktitle}{Encyclopedia of Condensed Matter Physics (Second
  Edition)}. \bibinfo{publisher}{Academic Press}, \bibinfo{address}{Oxford},
  pp. \bibinfo{pages}{693--701}.
\newblock \DOIprefix\doi{10.1016/B978-0-323-90800-9.00079-2}.
\bibitem[{Burnett et~al.(2017)Burnett, Sagar, Kennedy, Warburton and
  Fenton}]{BURN17}
\bibinfo{author}{Burnett, J.}, \bibinfo{author}{Sagar, J.},
  \bibinfo{author}{Kennedy, O.W.}, \bibinfo{author}{Warburton, P.A.},
  \bibinfo{author}{Fenton, J.C.}, \bibinfo{year}{2017}.
\newblock \bibinfo{title}{Low-loss superconducting nanowire circuits using a
  neon focused ion beam}.
\newblock \bibinfo{journal}{Phys. Rev. Appl.} \bibinfo{volume}{8},
  \bibinfo{pages}{014039}.
\newblock \DOIprefix\doi{10.1103/physrevapplied.8.014039}.
\bibitem[{Buzdin and Feinberg(1996)}]{BUZD96b}
\bibinfo{author}{Buzdin, A.}, \bibinfo{author}{Feinberg, D.},
  \bibinfo{year}{1996}.
\newblock \bibinfo{title}{Electromagnetic pinning of vortices by
  non-superconducting defects and their influence on screening}.
\newblock \bibinfo{journal}{Physica C} \bibinfo{volume}{256},
  \bibinfo{pages}{303--311}.
\newblock \DOIprefix\doi{10.1016/0921-4534(95)00664-8}.
\bibitem[{Buzdin(1993)}]{BUZD93}
\bibinfo{author}{Buzdin, A.I.}, \bibinfo{year}{1993}.
\newblock \bibinfo{title}{Multiple-quanta vortices at columnar defects}.
\newblock \bibinfo{journal}{Phys. Rev. B} \bibinfo{volume}{47},
  \bibinfo{pages}{11416}.
\newblock \DOIprefix\doi{10.1103/PhysRevB.47.11416}.
\bibitem[{Castellanos et~al.(1997)Castellanos, W\"ordenweber, Ockenfuss, v.d.
  Hart and Keck}]{CAST97}
\bibinfo{author}{Castellanos, A.}, \bibinfo{author}{W\"ordenweber, R.},
  \bibinfo{author}{Ockenfuss, G.}, \bibinfo{author}{v.d. Hart, A.},
  \bibinfo{author}{Keck, K.}, \bibinfo{year}{1997}.
\newblock \bibinfo{title}{Preparation of regular arrays of antidots in
  {YBa$_2$Cu$_3$O$_7$} thin films and observation of vortex lattice matching
  effects}.
\newblock \bibinfo{journal}{Appl. Phys. Lett.} \bibinfo{volume}{71},
  \bibinfo{pages}{962--964}.
\newblock \DOIprefix\doi{10.1063/1.119701}.
\bibitem[{Civale(1997)}]{CIVA97R}
\bibinfo{author}{Civale, L.}, \bibinfo{year}{1997}.
\newblock \bibinfo{title}{Vortex pinning and creep in high-temperature
  superconductors with columnar defects}.
\newblock \bibinfo{journal}{Supercond. Sci. Technol.} \bibinfo{volume}{10},
  \bibinfo{pages}{A11--A28}.
\newblock \DOIprefix\doi{10.1088/0953-2048/10/7a/003}.
\bibitem[{Cole et~al.(2006)Cole, Bending, Savel'ev, Grigorenko, Tamegai and
  Nori}]{COLE06}
\bibinfo{author}{Cole, D.}, \bibinfo{author}{Bending, S.},
  \bibinfo{author}{Savel'ev, S.}, \bibinfo{author}{Grigorenko, A.},
  \bibinfo{author}{Tamegai, T.}, \bibinfo{author}{Nori, F.},
  \bibinfo{year}{2006}.
\newblock \bibinfo{title}{Ratchet without spatial asymmetry for controlling the
  motion of magnetic flux quanta using time-asymmetric drives}.
\newblock \bibinfo{journal}{Nat. Mater.} \bibinfo{volume}{5},
  \bibinfo{pages}{305--311}.
\newblock \DOIprefix\doi{10.1038/nmat1608}.
\bibitem[{Collienne et~al.(2021)Collienne, Raes, Keijers, Linek, Koelle,
  Kleiner, Kramer, {Van~de~Vondel} and Silhanek}]{COLL21}
\bibinfo{author}{Collienne, S.}, \bibinfo{author}{Raes, B.},
  \bibinfo{author}{Keijers, W.}, \bibinfo{author}{Linek, J.},
  \bibinfo{author}{Koelle, D.}, \bibinfo{author}{Kleiner, R.},
  \bibinfo{author}{Kramer, R.B.G.}, \bibinfo{author}{{Van~de~Vondel}, J.},
  \bibinfo{author}{Silhanek, A.V.}, \bibinfo{year}{2021}.
\newblock \bibinfo{title}{Nb-based nanoscale superconducting quantum
  interference devices tuned by electroannealing}.
\newblock \bibinfo{journal}{Phys. Rev. Appl.} \bibinfo{volume}{15},
  \bibinfo{pages}{034016}.
\newblock \DOIprefix\doi{10.1103/physrevapplied.15.034016}.
\bibitem[{C\'ordoba(2024)}]{CORD24}
\bibinfo{author}{C\'ordoba, R.}, \bibinfo{year}{2024}.
\newblock \bibinfo{title}{Additive nanofabrication using focused ion and
  electron beams}, in: \bibinfo{editor}{Chakraborty, T.} (Ed.),
  \bibinfo{booktitle}{Encyclopedia of Condensed Matter Physics (Second
  Edition)}. \bibinfo{publisher}{Academic Press}, \bibinfo{address}{Oxford},
  pp. \bibinfo{pages}{448--464}.
\newblock \DOIprefix\doi{10.1016/B978-0-323-90800-9.00035-4}.
\bibitem[{Cuppens et~al.(2011)Cuppens, Ataklti, Gillijns, Van~de Vondel,
  Moshchalkov and Silhanek}]{CUPP11}
\bibinfo{author}{Cuppens, J.}, \bibinfo{author}{Ataklti, G.W.},
  \bibinfo{author}{Gillijns, W.}, \bibinfo{author}{Van~de Vondel, J.},
  \bibinfo{author}{Moshchalkov, V.V.}, \bibinfo{author}{Silhanek, A.V.},
  \bibinfo{year}{2011}.
\newblock \bibinfo{title}{Vortex dynamics in a superconducting film with a
  kagome and a honeycomb pinning landscape}.
\newblock \bibinfo{journal}{J. Supercond. Novel Magn.} \bibinfo{volume}{24},
  \bibinfo{pages}{7--11}.
\newblock \DOIprefix\doi{10.1007/s10948-010-0893-7}.
\bibitem[{Cybart et~al.(2009)Cybart, Anton, Wu, Clarke and Dynes}]{CYBA09}
\bibinfo{author}{Cybart, S.A.}, \bibinfo{author}{Anton, S.M.},
  \bibinfo{author}{Wu, S.M.}, \bibinfo{author}{Clarke, J.},
  \bibinfo{author}{Dynes, R.C.}, \bibinfo{year}{2009}.
\newblock \bibinfo{title}{Very large scale integration of nanopatterned
  {YBa}$_2${Cu}$_3${O}$_{7-\delta}$ {J}osephson junctions in a two-dimensional
  array}.
\newblock \bibinfo{journal}{Nano Lett.} \bibinfo{volume}{9},
  \bibinfo{pages}{3581--3585}.
\newblock \DOIprefix\doi{10.1021/nl901785j}.
\bibitem[{Cybart et~al.(2014)Cybart, Cho, Wong, Glyantsev, Huh, Yung, Moeckly,
  Beeman, Ulin-Avila, Wu and Dynes}]{CYBA14}
\bibinfo{author}{Cybart, S.A.}, \bibinfo{author}{Cho, E.Y.},
  \bibinfo{author}{Wong, T.J.}, \bibinfo{author}{Glyantsev, V.N.},
  \bibinfo{author}{Huh, J.U.}, \bibinfo{author}{Yung, C.S.},
  \bibinfo{author}{Moeckly, B.H.}, \bibinfo{author}{Beeman, J.W.},
  \bibinfo{author}{Ulin-Avila, E.}, \bibinfo{author}{Wu, S.M.},
  \bibinfo{author}{Dynes, R.C.}, \bibinfo{year}{2014}.
\newblock \bibinfo{title}{Large voltage modulation in magnetic field sensors
  from two-dimensional arrays of {Y-Ba-Cu-O} nano {J}osephson junctions}.
\newblock \bibinfo{journal}{Appl. Phys. Lett.} \bibinfo{volume}{104},
  \bibinfo{pages}{062601}.
\newblock \DOIprefix\doi{10.1063/1.4865216}.
\bibitem[{Cybart et~al.(2015)Cybart, Cho, Wong, Wehlin, Ma, Huynh and
  Dynes}]{CYBA15}
\bibinfo{author}{Cybart, S.A.}, \bibinfo{author}{Cho, E.Y.},
  \bibinfo{author}{Wong, T.J.}, \bibinfo{author}{Wehlin, B.H.},
  \bibinfo{author}{Ma, M.K.}, \bibinfo{author}{Huynh, C.},
  \bibinfo{author}{Dynes, R.C.}, \bibinfo{year}{2015}.
\newblock \bibinfo{title}{Nano {J}osephson superconducting tunnel junctions in
  {YBa}$_2${Cu}$_3${O}$_{7-\delta}$ directly patterned with a focused helium
  ion beam}.
\newblock \bibinfo{journal}{Nat. Nanotechnol.} \bibinfo{volume}{10},
  \bibinfo{pages}{598}.
\newblock \DOIprefix\doi{10.1038/nnano.2015.76}.
\bibitem[{Dobrovolskiy(2024)}]{DOBR24R}
\bibinfo{author}{Dobrovolskiy, O.V.}, \bibinfo{year}{2024}.
\newblock \bibinfo{title}{Fast dynamics of vortices in superconductors}, in:
  \bibinfo{editor}{Chakraborty, T.} (Ed.), \bibinfo{booktitle}{Encyclopedia of
  Condensed Matter Physics (Second Edition)}. \bibinfo{publisher}{Academic
  Press}, \bibinfo{address}{Oxford}, pp. \bibinfo{pages}{735--754}.
\newblock \DOIprefix\doi{10.1016/B978-0-323-90800-9.00015-9}.
\bibitem[{Dobrovolskiy et~al.(2016)Dobrovolskiy, Hanefeld, Z\"orb, Huth and
  Shklovskij}]{DOBR16}
\bibinfo{author}{Dobrovolskiy, O.V.}, \bibinfo{author}{Hanefeld, M.},
  \bibinfo{author}{Z\"orb, M.}, \bibinfo{author}{Huth, M.},
  \bibinfo{author}{Shklovskij, V.A.}, \bibinfo{year}{2016}.
\newblock \bibinfo{title}{Interplay of flux guiding and {H}all effect in {Nb}
  films with nanogrooves}.
\newblock \bibinfo{journal}{Supercond. Sci. Technol.} \bibinfo{volume}{29},
  \bibinfo{pages}{065009}.
\newblock \DOIprefix\doi{10.1088/0953-2048/29/6/065009}.
\bibitem[{Durrell et~al.(2000)Durrell, Gibson, Barber, Evetts, Rössler,
  Pedarnig and Bäuerle}]{DURR00}
\bibinfo{author}{Durrell, J.H.}, \bibinfo{author}{Gibson, G.},
  \bibinfo{author}{Barber, Z.H.}, \bibinfo{author}{Evetts, J.E.},
  \bibinfo{author}{Rössler, R.}, \bibinfo{author}{Pedarnig, J.D.},
  \bibinfo{author}{Bäuerle, D.}, \bibinfo{year}{2000}.
\newblock \bibinfo{title}{Dependence of critical current on field angle in
  off-c-axis grown {Bi$_2$Sr$_2$CaCu$_2$O$_8$} film}.
\newblock \bibinfo{journal}{Appl. Phys. Lett.} \bibinfo{volume}{77},
  \bibinfo{pages}{1686--8}.
\newblock \DOIprefix\doi{10.1063/1.1310174}.
\bibitem[{Emmrich et~al.(2016)Emmrich, Beyer, Nadzeyka, Bauerdick, Meyer,
  Kotakoski and G\"olzh\"auser}]{EMMR16}
\bibinfo{author}{Emmrich, D.}, \bibinfo{author}{Beyer, A.},
  \bibinfo{author}{Nadzeyka, A.}, \bibinfo{author}{Bauerdick, S.},
  \bibinfo{author}{Meyer, J.C.}, \bibinfo{author}{Kotakoski, J.},
  \bibinfo{author}{G\"olzh\"auser, A.}, \bibinfo{year}{2016}.
\newblock \bibinfo{title}{Nanopore fabrication and characterization by helium
  ion microscopy}.
\newblock \bibinfo{journal}{Appl. Phys. Lett.} \bibinfo{volume}{108},
  \bibinfo{pages}{163103}.
\newblock \DOIprefix\doi{10.1063/1.4947277}.
\bibitem[{Fisher et~al.(1991)Fisher, Fisher and Huse}]{FISH91}
\bibinfo{author}{Fisher, D.S.}, \bibinfo{author}{Fisher, M.P.A.},
  \bibinfo{author}{Huse, D.A.}, \bibinfo{year}{1991}.
\newblock \bibinfo{title}{Thermal fluctuations, quenched disorder, phase
  transitions, and transport in type-{II} superconductors}.
\newblock \bibinfo{journal}{Phys. Rev.} \bibinfo{volume}{43},
  \bibinfo{pages}{130--159}.
\newblock \DOIprefix\doi{10.1103/physrevb.43.130}.
\bibitem[{Fomin(2020)}]{FOMI20M}
\bibinfo{author}{Fomin, V.M.}, \bibinfo{year}{2020}.
\newblock \bibinfo{title}{Self-rolled Micro- and Nanoarchitectures}.
\newblock \bibinfo{publisher}{De Gruyter}, \bibinfo{address}{Berlin/Boston}.
\newblock \DOIprefix\doi{10.1515/9783110575576}.
\bibitem[{Gol'tsman et~al.(2001)Gol'tsman, Okunev, Chulkova, Lipatov, Semenov,
  Smirnov, Voronov, Dzardanov, Williams and Sobolewski}]{GOLT01}
\bibinfo{author}{Gol'tsman, G.N.}, \bibinfo{author}{Okunev, O.},
  \bibinfo{author}{Chulkova, G.}, \bibinfo{author}{Lipatov, A.},
  \bibinfo{author}{Semenov, A.}, \bibinfo{author}{Smirnov, K.},
  \bibinfo{author}{Voronov, B.}, \bibinfo{author}{Dzardanov, A.},
  \bibinfo{author}{Williams, C.}, \bibinfo{author}{Sobolewski, R.},
  \bibinfo{year}{2001}.
\newblock \bibinfo{title}{Picosecond superconducting single-photon optical
  detector}.
\newblock \bibinfo{journal}{Appl. Phys. Lett.} \bibinfo{volume}{79},
  \bibinfo{pages}{705--707}.
\newblock \DOIprefix\doi{10.1063/1.1388868}.
\bibitem[{Gozar et~al.(2017)Gozar, Litombe, Hoffman and
  Bo{\v{z}}ovi{\'{c}}}]{GOZA17}
\bibinfo{author}{Gozar, A.}, \bibinfo{author}{Litombe, N.E.},
  \bibinfo{author}{Hoffman, J.E.}, \bibinfo{author}{Bo{\v{z}}ovi{\'{c}}, I.},
  \bibinfo{year}{2017}.
\newblock \bibinfo{title}{Optical nanoscopy of high {$T_c$} cuprate
  nanoconstriction devices patterned by helium ion beams}.
\newblock \bibinfo{journal}{Nano Lett.} \bibinfo{volume}{17},
  \bibinfo{pages}{1582--1586}.
\newblock \DOIprefix\doi{10.1021/acs.nanolett.6b04729}.
\bibitem[{Gray et~al.(2022)Gray, Rushton and Murphy}]{GRAY22}
\bibinfo{author}{Gray, R.L.}, \bibinfo{author}{Rushton, M.J.D.},
  \bibinfo{author}{Murphy, S.T.}, \bibinfo{year}{2022}.
\newblock \bibinfo{title}{Molecular dynamics simulations of radiation damage in
  {YBa$_{2}$Cu$_{3}$O$_{7}$}}.
\newblock \bibinfo{journal}{Supercond. Sci. Technol.} \bibinfo{volume}{35},
  \bibinfo{pages}{035010}.
\newblock \DOIprefix\doi{10.1088/1361-6668/ac47dc}.
\bibitem[{Haag et~al.(2014)Haag, Zechner, Lang, Dosmailov, Bodea and
  Pedarnig}]{HAAG14}
\bibinfo{author}{Haag, L.T.}, \bibinfo{author}{Zechner, G.},
  \bibinfo{author}{Lang, W.}, \bibinfo{author}{Dosmailov, M.},
  \bibinfo{author}{Bodea, M.A.}, \bibinfo{author}{Pedarnig, J.D.},
  \bibinfo{year}{2014}.
\newblock \bibinfo{title}{Strong vortex matching effects in {YBCO} films with
  periodic modulations of the superconducting order parameter fabricated by
  masked ion irradiation}.
\newblock \bibinfo{journal}{Physica C} \bibinfo{volume}{503},
  \bibinfo{pages}{75--81}.
\newblock \DOIprefix\doi{10.1016/j.physc.2014.03.032}.
\bibitem[{Haage et~al.(1997)Haage, Zegenhagen, Li, Habermeier, Cardona,
  Warthmann, Forkl and Kronm\"uller}]{HAAG97}
\bibinfo{author}{Haage, T.}, \bibinfo{author}{Zegenhagen, J.},
  \bibinfo{author}{Li, J.Q.}, \bibinfo{author}{Habermeier, H.U.},
  \bibinfo{author}{Cardona, M.}, \bibinfo{author}{Warthmann, J.R.},
  \bibinfo{author}{Forkl, A.}, \bibinfo{author}{Kronm\"uller, H.},
  \bibinfo{year}{1997}.
\newblock \bibinfo{title}{Transport properties and flux pinning by
  self-organization in {YBa$_{2}$Cu$_{3}$O$_{7-\delta}$} films on vicinal
  {SrTiO$_{3}$} (001)}.
\newblock \bibinfo{journal}{Phys. Rev. B} \bibinfo{volume}{56},
  \bibinfo{pages}{8404--8418}.
\newblock \DOIprefix\doi{10.1103/PhysRevB.56.8404}.
\bibitem[{Hadfield(2009)}]{HADF09R}
\bibinfo{author}{Hadfield, R.H.}, \bibinfo{year}{2009}.
\newblock \bibinfo{title}{Single-photon detectors for optical quantum
  information applications}.
\newblock \bibinfo{journal}{Nat. Photon.} \bibinfo{volume}{3},
  \bibinfo{pages}{696--705}.
\newblock \DOIprefix\doi{10.1038/nphoton.2009.230}.
\bibitem[{H\"anggi and Marchesoni(2009)}]{HANG09R}
\bibinfo{author}{H\"anggi, P.}, \bibinfo{author}{Marchesoni, F.},
  \bibinfo{year}{2009}.
\newblock \bibinfo{title}{Artificial {B}rownian motors: {C}ontrolling transport
  on the nanoscale}.
\newblock \bibinfo{journal}{Rev. Mod. Phys.} \bibinfo{volume}{81},
  \bibinfo{pages}{387--442}.
\newblock \DOIprefix\doi{10.1103/RevModPhys.81.387}.
\bibitem[{Harada et~al.(1996)Harada, Kamimura, Kasai, Matsuda, Tonomura and
  Moshchalkov}]{HARA96b}
\bibinfo{author}{Harada, K.}, \bibinfo{author}{Kamimura, O.},
  \bibinfo{author}{Kasai, H.}, \bibinfo{author}{Matsuda, T.},
  \bibinfo{author}{Tonomura, A.}, \bibinfo{author}{Moshchalkov, V.V.},
  \bibinfo{year}{1996}.
\newblock \bibinfo{title}{Direct observation of vortex dynamics in
  superconducting films with regular arrays of defects}.
\newblock \bibinfo{journal}{Science} \bibinfo{volume}{274},
  \bibinfo{pages}{1167--1170}.
\newblock \DOIprefix\doi{10.1126/science.274.5290.1167}.
\bibitem[{Hastings et~al.(2003)Hastings, Olson~Reichhardt and
  Reichhardt}]{HAST03}
\bibinfo{author}{Hastings, M.B.}, \bibinfo{author}{Olson~Reichhardt, C.J.},
  \bibinfo{author}{Reichhardt, C.}, \bibinfo{year}{2003}.
\newblock \bibinfo{title}{Ratchet cellular automata}.
\newblock \bibinfo{journal}{Phys. Rev. Lett.} \bibinfo{volume}{90},
  \bibinfo{pages}{247004}.
\newblock \DOIprefix\doi{10.1103/physrevlett.90.247004}.
\bibitem[{Heine et~al.(2021)Heine, Lang, Rössler and Pedarnig}]{HEIN21}
\bibinfo{author}{Heine, G.}, \bibinfo{author}{Lang, W.},
  \bibinfo{author}{Rössler, R.}, \bibinfo{author}{Pedarnig, J.D.},
  \bibinfo{year}{2021}.
\newblock \bibinfo{title}{Anisotropy of the in-plane and out-of-plane
  resistivity and the {H}all effect in the normal state of vicinal-grown
  {YBa$_{2}$Cu$_{3}$O$_{7-\delta}$} thin films}.
\newblock \bibinfo{journal}{Nanomaterials} \bibinfo{volume}{11},
  \bibinfo{pages}{675}.
\newblock \DOIprefix\doi{10.3390/nano11030675}.
\bibitem[{Hlawacek and G\"olzh\"auser(2016)}]{HLAW16M}
\bibinfo{editor}{Hlawacek, G.}, \bibinfo{editor}{G\"olzh\"auser, A.} (Eds.),
  \bibinfo{year}{2016}.
\newblock \bibinfo{title}{Helium Ion Microscopy}.
\newblock \bibinfo{publisher}{Springer International Publishing},
  \bibinfo{address}{Switzerland}.
\newblock \DOIprefix\doi{10.1007/978-3-319-41990-9}.
\bibitem[{Kahlmann et~al.(1998)Kahlmann, Engelhardt, Schubert, Zander, Buchal
  and Hollkott}]{KAHL98}
\bibinfo{author}{Kahlmann, F.}, \bibinfo{author}{Engelhardt, A.},
  \bibinfo{author}{Schubert, J.}, \bibinfo{author}{Zander, W.},
  \bibinfo{author}{Buchal, C.}, \bibinfo{author}{Hollkott, J.},
  \bibinfo{year}{1998}.
\newblock \bibinfo{title}{Superconductor-normal-superconductor {J}osephson
  junctions fabricated by oxygen implantation into
  {YBa$_2$Cu$_3$O$_{7-\delta}$}}.
\newblock \bibinfo{journal}{Appl. Phys. Lett.} \bibinfo{volume}{73},
  \bibinfo{pages}{2354--6}.
\newblock \DOIprefix\doi{10.1063/1.122459}.
\bibitem[{Kang et~al.(2002)Kang, Burnell, Lloyd, Speaks, Peng, Jeynes, Webb,
  Yun, Moon, Oh, Tarte, Moore and Blamire}]{KANG02a}
\bibinfo{author}{Kang, D.J.}, \bibinfo{author}{Burnell, G.},
  \bibinfo{author}{Lloyd, S.J.}, \bibinfo{author}{Speaks, R.S.},
  \bibinfo{author}{Peng, N.H.}, \bibinfo{author}{Jeynes, C.},
  \bibinfo{author}{Webb, R.}, \bibinfo{author}{Yun, J.H.},
  \bibinfo{author}{Moon, S.H.}, \bibinfo{author}{Oh, B.},
  \bibinfo{author}{Tarte, E.J.}, \bibinfo{author}{Moore, D.F.},
  \bibinfo{author}{Blamire, M.G.}, \bibinfo{year}{2002}.
\newblock \bibinfo{title}{Realization and properties of
  {YBa$_2$Cu$_3$O$_{7-\delta}$} {J}osephson junctions by metal masked ion
  damage technique}.
\newblock \bibinfo{journal}{Appl. Phys. Lett.} \bibinfo{volume}{80},
  \bibinfo{pages}{814--816}.
\newblock \DOIprefix\doi{10.1063/1.1446998}.
\bibitem[{Karrer et~al.(2023)Karrer, Aichner, Wurster, Kleiner, Goldobin,
  Koelle and Lang}]{KARR22P}
\bibinfo{author}{Karrer, M.}, \bibinfo{author}{Aichner, B.},
  \bibinfo{author}{Wurster, K.}, \bibinfo{author}{Kleiner, R.},
  \bibinfo{author}{Goldobin, E.}, \bibinfo{author}{Koelle, D.},
  \bibinfo{author}{Lang, W.}, \bibinfo{year}{2023}.
\newblock \bibinfo{title}{High-magnetic-field commensurability effects in
  ultradense pinning lattices fabricated by focused {He}-ion beam}.
\newblock \bibinfo{note}{In preparation}.
\bibitem[{Kasaei et~al.(2018)Kasaei, Melbourne, Manichev, Feldman, Gustafsson,
  Chen, Xi and Davidson}]{KASA18}
\bibinfo{author}{Kasaei, L.}, \bibinfo{author}{Melbourne, T.},
  \bibinfo{author}{Manichev, V.}, \bibinfo{author}{Feldman, L.C.},
  \bibinfo{author}{Gustafsson, T.}, \bibinfo{author}{Chen, K.},
  \bibinfo{author}{Xi, X.X.}, \bibinfo{author}{Davidson, B.A.},
  \bibinfo{year}{2018}.
\newblock \bibinfo{title}{{MgB$_2$} {J}osephson junctions produced by focused
  helium ion beam irradiation}.
\newblock \bibinfo{journal}{AIP Adv.} \bibinfo{volume}{8},
  \bibinfo{pages}{075020}.
\newblock \DOIprefix\doi{10.1063/1.5030751}.
\bibitem[{Katz et~al.(1998)Katz, Sun, Woods and Dynes}]{KATZ98}
\bibinfo{author}{Katz, A.S.}, \bibinfo{author}{Sun, A.G.},
  \bibinfo{author}{Woods, S.I.}, \bibinfo{author}{Dynes, R.C.},
  \bibinfo{year}{1998}.
\newblock \bibinfo{title}{Planar thin film {YBa$_2$Cu$_3$O$_{7- \delta}$}
  {J}osephson junctions via nanolithography and ion damage}.
\newblock \bibinfo{journal}{Appl. Phys. Lett.} \bibinfo{volume}{72},
  \bibinfo{pages}{2032--2034}.
\newblock \DOIprefix\doi{10.1063/1.121255}.
\bibitem[{Katz et~al.(2000)Katz, Woods and Dynes}]{KATZ00}
\bibinfo{author}{Katz, A.S.}, \bibinfo{author}{Woods, S.I.},
  \bibinfo{author}{Dynes, R.C.}, \bibinfo{year}{2000}.
\newblock \bibinfo{title}{Transport properties of high-{$T_c$} planar
  {J}osephson junctions fabricated by nanolithography and ion implantation}.
\newblock \bibinfo{journal}{J. Appl. Phys.} \bibinfo{volume}{87},
  \bibinfo{pages}{2978--83}.
\newblock \DOIprefix\doi{10.1063/1.372286}.
\bibitem[{Kemmler et~al.(2006)Kemmler, G{\"u}rlich, Sterck, P{\"o}hler,
  Neuhaus, Siegel, Kleiner and Koelle}]{KEMM06}
\bibinfo{author}{Kemmler, M.}, \bibinfo{author}{G{\"u}rlich, C.},
  \bibinfo{author}{Sterck, A.}, \bibinfo{author}{P{\"o}hler, H.},
  \bibinfo{author}{Neuhaus, M.}, \bibinfo{author}{Siegel, M.},
  \bibinfo{author}{Kleiner, R.}, \bibinfo{author}{Koelle, D.},
  \bibinfo{year}{2006}.
\newblock \bibinfo{title}{Commensurability effects in superconducting {Nb}
  films with quasiperiodic pinning arrays}.
\newblock \bibinfo{journal}{Phys. Rev. Lett.} \bibinfo{volume}{97},
  \bibinfo{pages}{147003}.
\newblock \DOIprefix\doi{10.1103/physrevlett.97.147003}.
\bibitem[{Koelle et~al.(1999)Koelle, Kleiner, Ludwig, Dantsker and
  Clarke}]{KOEL99R}
\bibinfo{author}{Koelle, D.}, \bibinfo{author}{Kleiner, R.},
  \bibinfo{author}{Ludwig, F.}, \bibinfo{author}{Dantsker, E.},
  \bibinfo{author}{Clarke, J.}, \bibinfo{year}{1999}.
\newblock \bibinfo{title}{High-transition-temperature superconducting quantum
  interference devices}.
\newblock \bibinfo{journal}{Rev. Mod. Phys.} \bibinfo{volume}{71},
  \bibinfo{pages}{631--686}.
\newblock \DOIprefix\doi{10.1103/revmodphys.71.631}.
\bibitem[{Kramer et~al.(2009)Kramer, Silhanek, {Van de Vondel}, Raes and
  Moshchalkov}]{KRAM09}
\bibinfo{author}{Kramer, R.B.G.}, \bibinfo{author}{Silhanek, A.V.},
  \bibinfo{author}{{Van de Vondel}, J.}, \bibinfo{author}{Raes, B.},
  \bibinfo{author}{Moshchalkov, V.V.}, \bibinfo{year}{2009}.
\newblock \bibinfo{title}{Symmetry-induced giant vortex state in a
  superconducting {Pb} film with a fivefold {P}enrose array of magnetic pinning
  centers}.
\newblock \bibinfo{journal}{Phys. Rev. Lett.} \bibinfo{volume}{103},
  \bibinfo{pages}{067007}.
\newblock \DOIprefix\doi{10.1103/physrevlett.103.067007}.
\bibitem[{Laguna et~al.(2001)Laguna, Balseiro, Dom\'{\i}nguez and
  Nori}]{LAGU01}
\bibinfo{author}{Laguna, M.F.}, \bibinfo{author}{Balseiro, C.A.},
  \bibinfo{author}{Dom\'{\i}nguez, D.}, \bibinfo{author}{Nori, F.},
  \bibinfo{year}{2001}.
\newblock \bibinfo{title}{Vortex structure and dynamics in kagom\'e and
  triangular pinning potentials}.
\newblock \bibinfo{journal}{Phys. Rev. B} \bibinfo{volume}{64},
  \bibinfo{pages}{104505}.
\newblock \DOIprefix\doi{10.1103/PhysRevB.64.104505}.
\bibitem[{Lang et~al.(2006)Lang, Dineva, Marksteiner, Enzenhofer, Siraj,
  Peruzzi, Pedarnig, B\"auerle, Korntner, Cekan, Platzgummer and
  Loeschner}]{LANG06a}
\bibinfo{author}{Lang, W.}, \bibinfo{author}{Dineva, M.},
  \bibinfo{author}{Marksteiner, M.}, \bibinfo{author}{Enzenhofer, T.},
  \bibinfo{author}{Siraj, K.}, \bibinfo{author}{Peruzzi, M.},
  \bibinfo{author}{Pedarnig, J.D.}, \bibinfo{author}{B\"auerle, D.},
  \bibinfo{author}{Korntner, R.}, \bibinfo{author}{Cekan, E.},
  \bibinfo{author}{Platzgummer, E.}, \bibinfo{author}{Loeschner, H.},
  \bibinfo{year}{2006}.
\newblock \bibinfo{title}{Ion-beam direct-structuring of high-temperature
  superconductors}.
\newblock \bibinfo{journal}{Microelectron. Eng.} \bibinfo{volume}{83},
  \bibinfo{pages}{1495--1498}.
\newblock \DOIprefix\doi{10.1016/j.mee.2006.01.091}.
\bibitem[{Lang and Pedarnig(2010)}]{LANG10R}
\bibinfo{author}{Lang, W.}, \bibinfo{author}{Pedarnig, J.D.},
  \bibinfo{year}{2010}.
\newblock \bibinfo{title}{Ion irradiation of high-temperature su-perconductors
  and its application for nanopatterning}, in: \bibinfo{editor}{Moshchalkov,
  V.V.}, \bibinfo{editor}{Wördenweber, R.}, \bibinfo{editor}{Lang, W.} (Eds.),
  \bibinfo{booktitle}{Nanoscience and Engineering in Superconductivity}.
  \bibinfo{publisher}{Springer}, \bibinfo{address}{Heidelberg}, pp.
  \bibinfo{pages}{81--104}.
\newblock \DOIprefix\doi{10.1007/978-3-642-15137-8}.
\bibitem[{Latimer et~al.(2012)Latimer, Berdiyorov, Xiao, Kwok and
  Peeters}]{LATI12}
\bibinfo{author}{Latimer, M.L.}, \bibinfo{author}{Berdiyorov, G.R.},
  \bibinfo{author}{Xiao, Z.L.}, \bibinfo{author}{Kwok, W.K.},
  \bibinfo{author}{Peeters, F.M.}, \bibinfo{year}{2012}.
\newblock \bibinfo{title}{Vortex interaction enhanced saturation number and
  caging effect in a superconducting film with a honeycomb array of nanoscale
  holes}.
\newblock \bibinfo{journal}{Phys. Rev. B} \bibinfo{volume}{85},
  \bibinfo{pages}{012505}.
\newblock \DOIprefix\doi{10.1103/PhysRevB.85.012505}.
\bibitem[{Latimer et~al.(2013)Latimer, Berdiyorov, Xiao, Peeters and
  Kwok}]{LATI13}
\bibinfo{author}{Latimer, M.L.}, \bibinfo{author}{Berdiyorov, G.R.},
  \bibinfo{author}{Xiao, Z.L.}, \bibinfo{author}{Peeters, F.M.},
  \bibinfo{author}{Kwok, W.K.}, \bibinfo{year}{2013}.
\newblock \bibinfo{title}{Realization of artificial ice systems for magnetic
  vortices in a superconducting {MoGe} thin film with patterned
  nanostructures}.
\newblock \bibinfo{journal}{Phys. Rev. Lett.} \bibinfo{volume}{111},
  \bibinfo{pages}{067001}.
\newblock \DOIprefix\doi{10.1103/PhysRevLett.111.067001}.
\bibitem[{Laviano et~al.(2010)Laviano, Ghigo, Mezzetti, Hollmann and
  W\"ordenweber}]{LAVI10}
\bibinfo{author}{Laviano, F.}, \bibinfo{author}{Ghigo, G.},
  \bibinfo{author}{Mezzetti, E.}, \bibinfo{author}{Hollmann, E.},
  \bibinfo{author}{W\"ordenweber, R.}, \bibinfo{year}{2010}.
\newblock \bibinfo{title}{Control of the vortex flow in microchannel arrays
  produced in {YBCO} films by heavy-ion lithography}.
\newblock \bibinfo{journal}{Physica C} \bibinfo{volume}{470},
  \bibinfo{pages}{844--847}.
\newblock \DOIprefix\doi{10.1016/j.physc.2010.02.052}.
\bibitem[{Lee et~al.(1999)Lee, Jank\'o, Der\'enyi and Barab\'asi}]{LEE99}
\bibinfo{author}{Lee, C.S.}, \bibinfo{author}{Jank\'o, B.},
  \bibinfo{author}{Der\'enyi, I.}, \bibinfo{author}{Barab\'asi, A.L.},
  \bibinfo{year}{1999}.
\newblock \bibinfo{title}{Reducing vortex density in superconductors using the
  `ratchet effect'}.
\newblock \bibinfo{journal}{Nature} \bibinfo{volume}{400},
  \bibinfo{pages}{337--340}.
\newblock \DOIprefix\doi{10.1038/22485}.
\bibitem[{LeFebvre et~al.(2019)LeFebvre, Cho, Li, Pratt and Cybart}]{LEFE19}
\bibinfo{author}{LeFebvre, J.C.}, \bibinfo{author}{Cho, E.},
  \bibinfo{author}{Li, H.}, \bibinfo{author}{Pratt, K.},
  \bibinfo{author}{Cybart, S.A.}, \bibinfo{year}{2019}.
\newblock \bibinfo{title}{Series arrays of planar long {J}osephson junctions
  for high dynamic range magnetic flux detection}.
\newblock \bibinfo{journal}{{AIP} Adv.} \bibinfo{volume}{9},
  \bibinfo{pages}{105215}.
\newblock \DOIprefix\doi{10.1063/1.5126035}.
\bibitem[{Lib{\'a}l et~al.(2009)Lib{\'a}l, Olson~Reichhardt and
  Reichhardt}]{LIBA09}
\bibinfo{author}{Lib{\'a}l, A.}, \bibinfo{author}{Olson~Reichhardt, C.J.},
  \bibinfo{author}{Reichhardt, C.}, \bibinfo{year}{2009}.
\newblock \bibinfo{title}{Creating artificial ice states using vortices in
  nanostructured superconductors}.
\newblock \bibinfo{journal}{Phys. Rev. Lett.} \bibinfo{volume}{102},
  \bibinfo{pages}{237004}.
\newblock \DOIprefix\doi{10.1103/physrevlett.102.237004}.
\bibitem[{Likharev(2012)}]{LIKH12R}
\bibinfo{author}{Likharev, K.K.}, \bibinfo{year}{2012}.
\newblock \bibinfo{title}{Superconductor digital electronics}.
\newblock \bibinfo{journal}{Physica C} \bibinfo{volume}{482},
  \bibinfo{pages}{6--18}.
\newblock \DOIprefix\doi{10.1016/j.physc.2012.05.016}.
\bibitem[{Lyu et~al.(2021)Lyu, Jiang, Wang, Xiao, Dong, Chen,
  Milo{\v{s}}evi{\'{c}}, Wang, Divan, Pearson, Wu, Peeters and Kwok}]{LYU21}
\bibinfo{author}{Lyu, Y.Y.}, \bibinfo{author}{Jiang, J.},
  \bibinfo{author}{Wang, Y.L.}, \bibinfo{author}{Xiao, Z.L.},
  \bibinfo{author}{Dong, S.}, \bibinfo{author}{Chen, Q.H.},
  \bibinfo{author}{Milo{\v{s}}evi{\'{c}}, M.V.}, \bibinfo{author}{Wang, H.},
  \bibinfo{author}{Divan, R.}, \bibinfo{author}{Pearson, J.E.},
  \bibinfo{author}{Wu, P.}, \bibinfo{author}{Peeters, F.M.},
  \bibinfo{author}{Kwok, W.K.}, \bibinfo{year}{2021}.
\newblock \bibinfo{title}{Superconducting diode effect via conformal-mapped
  nanoholes}.
\newblock \bibinfo{journal}{Nat. Commun.} \bibinfo{volume}{12},
  \bibinfo{pages}{2703}.
\newblock \DOIprefix\doi{10.1038/s41467-021-23077-0}.
\bibitem[{Marinkovi{\'{c}} et~al.(2020)Marinkovi{\'{c}},
  Fern{\'{a}}ndez-Rodr{\'{\i}}guez, Collienne, Alvarez, Melinte, Maiorov, Rius,
  Granados, Mestres, Palau and Silhanek}]{MARI20}
\bibinfo{author}{Marinkovi{\'{c}}, S.},
  \bibinfo{author}{Fern{\'{a}}ndez-Rodr{\'{\i}}guez, A.},
  \bibinfo{author}{Collienne, S.}, \bibinfo{author}{Alvarez, S.B.},
  \bibinfo{author}{Melinte, S.}, \bibinfo{author}{Maiorov, B.},
  \bibinfo{author}{Rius, G.}, \bibinfo{author}{Granados, X.},
  \bibinfo{author}{Mestres, N.}, \bibinfo{author}{Palau, A.},
  \bibinfo{author}{Silhanek, A.V.}, \bibinfo{year}{2020}.
\newblock \bibinfo{title}{Direct visualization of current-stimulated oxygen
  migration in {YBa$_{2}$Cu$_{3}$O$_{7-\delta}$} thin films}.
\newblock \bibinfo{journal}{{ACS} Nano} \bibinfo{volume}{14},
  \bibinfo{pages}{11765--11774}.
\newblock \DOIprefix\doi{10.1021/acsnano.0c04492}.
\bibitem[{Markowitsch et~al.(1997)Markowitsch, Stockinger, Lang, Bierleutgeb,
  Pedarnig and Bäuerle}]{MARK97}
\bibinfo{author}{Markowitsch, W.}, \bibinfo{author}{Stockinger, C.},
  \bibinfo{author}{Lang, W.}, \bibinfo{author}{Bierleutgeb, K.},
  \bibinfo{author}{Pedarnig, J.D.}, \bibinfo{author}{Bäuerle, D.},
  \bibinfo{year}{1997}.
\newblock \bibinfo{title}{Photoinduced enhancement of the c-axis conductivity
  in oxygen-deficient {YBa$_{2}$Cu$_{3}$O$_{7-\delta}$} thin films}.
\newblock \bibinfo{journal}{Appl. Phys. Lett.} \bibinfo{volume}{71},
  \bibinfo{pages}{1246--1248}.
\newblock \DOIprefix\doi{10.1063/1.119863}.
\bibitem[{Martin et~al.(1997)Martin, V\'elez, Nogu\'es and Schuller}]{MART97a}
\bibinfo{author}{Martin, J.I.}, \bibinfo{author}{V\'elez, M.},
  \bibinfo{author}{Nogu\'es, J.}, \bibinfo{author}{Schuller, I.K.},
  \bibinfo{year}{1997}.
\newblock \bibinfo{title}{Flux pinning in a superconductor by an array of
  submicrometer magnetic dots}.
\newblock \bibinfo{journal}{Phys. Rev. Lett.} \bibinfo{volume}{79},
  \bibinfo{pages}{1929--1932}.
\newblock \DOIprefix\doi{10.1103/PhysRevLett.79.1929}.
\bibitem[{Milo{\v{s}}evi{\'c} et~al.(2007)Milo{\v{s}}evi{\'c}, Berdiyorov and
  Peeters}]{MILO07}
\bibinfo{author}{Milo{\v{s}}evi{\'c}, M.V.}, \bibinfo{author}{Berdiyorov,
  G.R.}, \bibinfo{author}{Peeters, F.M.}, \bibinfo{year}{2007}.
\newblock \bibinfo{title}{Fluxonic cellular automata}.
\newblock \bibinfo{journal}{Appl. Phys. Lett.} \bibinfo{volume}{91},
  \bibinfo{pages}{212501}.
\newblock \DOIprefix\doi{10.1063/1.2813047}.
\bibitem[{Misko et~al.(2005)Misko, Savel'ev and Nori}]{MISK05}
\bibinfo{author}{Misko, V.}, \bibinfo{author}{Savel'ev, S.},
  \bibinfo{author}{Nori, F.}, \bibinfo{year}{2005}.
\newblock \bibinfo{title}{Critical currents in quasiperiodic pinning arrays:
  Chains and penrose lattices}.
\newblock \bibinfo{journal}{Phys. Rev. Lett.} \bibinfo{volume}{95},
  \bibinfo{pages}{177007}.
\newblock \DOIprefix\doi{10.1103/PhysRevLett.95.177007}.
\bibitem[{Misko et~al.(2010)Misko, Bothner, Kemmler, Kleiner, Koelle, Peeters
  and Nori}]{MISK10}
\bibinfo{author}{Misko, V.R.}, \bibinfo{author}{Bothner, D.},
  \bibinfo{author}{Kemmler, M.}, \bibinfo{author}{Kleiner, R.},
  \bibinfo{author}{Koelle, D.}, \bibinfo{author}{Peeters, F.M.},
  \bibinfo{author}{Nori, F.}, \bibinfo{year}{2010}.
\newblock \bibinfo{title}{Enhancing the critical current in quasiperiodic
  pinning arrays below and above the matching magnetic flux}.
\newblock \bibinfo{journal}{Phys. Rev. B} \bibinfo{volume}{82},
  \bibinfo{pages}{184512}.
\newblock \DOIprefix\doi{10.1103/PhysRevB.82.184512}.
\bibitem[{Misko and Nori(2012)}]{MISK12}
\bibinfo{author}{Misko, V.R.}, \bibinfo{author}{Nori, F.},
  \bibinfo{year}{2012}.
\newblock \bibinfo{title}{Magnetic flux pinning in superconductors with
  hyperbolic-tessellation arrays of pinning sites}.
\newblock \bibinfo{journal}{Phys. Rev. B} \bibinfo{volume}{85},
  \bibinfo{pages}{184506}.
\newblock \DOIprefix\doi{10.1103/PhysRevB.85.184506}.
\bibitem[{Moshchalkov et~al.(1996)Moshchalkov, Baert, Metlushko, Rosseel, Bael,
  Temst, Jonckheere and Bruynseraede}]{MOSH96a}
\bibinfo{author}{Moshchalkov, V.V.}, \bibinfo{author}{Baert, M.},
  \bibinfo{author}{Metlushko, V.V.}, \bibinfo{author}{Rosseel, E.},
  \bibinfo{author}{Bael, M.J.V.}, \bibinfo{author}{Temst, K.},
  \bibinfo{author}{Jonckheere, R.}, \bibinfo{author}{Bruynseraede, Y.},
  \bibinfo{year}{1996}.
\newblock \bibinfo{title}{Magnetization of multiple-quanta vortex lattices}.
\newblock \bibinfo{journal}{Phys. Rev. B} \bibinfo{volume}{54},
  \bibinfo{pages}{7385--7393}.
\newblock \DOIprefix\doi{10.1103/physrevb.54.7385}.
\bibitem[{Moshchalkov and Fritzsche(2011)}]{MOSH11M}
\bibinfo{author}{Moshchalkov, V.V.}, \bibinfo{author}{Fritzsche, J.},
  \bibinfo{year}{2011}.
\newblock \bibinfo{title}{Nanostructured superconductors}.
\newblock \bibinfo{publisher}{World Scientific}, \bibinfo{address}{Singapore}.
\newblock \DOIprefix\doi{10.1142/9789814343923}.
\bibitem[{M\"uller et~al.(2019)M\"uller, Karrer, Limberger, Becker,
  Schr\"oppel, Burkhardt, Kleiner, Goldobin and Koelle}]{MULL19}
\bibinfo{author}{M\"uller, B.}, \bibinfo{author}{Karrer, M.},
  \bibinfo{author}{Limberger, F.}, \bibinfo{author}{Becker, M.},
  \bibinfo{author}{Schr\"oppel, B.}, \bibinfo{author}{Burkhardt, C.J.},
  \bibinfo{author}{Kleiner, R.}, \bibinfo{author}{Goldobin, E.},
  \bibinfo{author}{Koelle, D.}, \bibinfo{year}{2019}.
\newblock \bibinfo{title}{Josephson junctions and {SQUIDs} created by focused
  helium-ion-beam irradiation of
  {${\mathrm{Y}\mathrm{Ba}}_{2}{\mathrm{Cu}}_{3}{\mathrm{O}}_{7}$}}.
\newblock \bibinfo{journal}{Phys. Rev. Applied} \bibinfo{volume}{11},
  \bibinfo{pages}{044082}.
\newblock \DOIprefix\doi{10.1103/PhysRevApplied.11.044082}.
\bibitem[{Nelson and Vinokur(1993)}]{NELS93}
\bibinfo{author}{Nelson, D.R.}, \bibinfo{author}{Vinokur, V.M.},
  \bibinfo{year}{1993}.
\newblock \bibinfo{title}{Boson localization and correlated pinning of
  superconducting vortex arrays}.
\newblock \bibinfo{journal}{Phys. Rev. B} \bibinfo{volume}{48},
  \bibinfo{pages}{13060--13097}.
\newblock \DOIprefix\doi{10.1103/physrevb.48.13060}.
\bibitem[{Parks and Little(1964)}]{PARK64}
\bibinfo{author}{Parks, R.D.}, \bibinfo{author}{Little, W.A.},
  \bibinfo{year}{1964}.
\newblock \bibinfo{title}{Fluxoid quantization in a multiply-connected
  superconductor}.
\newblock \bibinfo{journal}{Phys. Rev.} \bibinfo{volume}{133},
  \bibinfo{pages}{A97--A103}.
\newblock \DOIprefix\doi{10.1103/PhysRev.133.A97}.
\bibitem[{Pedarnig et~al.(2002)Pedarnig, R\"ossler, Delamare, Lang, B\"auerle,
  K\"ohler and Zandbergen}]{PEDA02}
\bibinfo{author}{Pedarnig, J.D.}, \bibinfo{author}{R\"ossler, R.},
  \bibinfo{author}{Delamare, M.P.}, \bibinfo{author}{Lang, W.},
  \bibinfo{author}{B\"auerle, D.}, \bibinfo{author}{K\"ohler, A.},
  \bibinfo{author}{Zandbergen, H.W.}, \bibinfo{year}{2002}.
\newblock \bibinfo{title}{Electrical properties, texture, and microstructure of
  vicinal {YBa$_2$Cu$_3$O$_{7-\delta}$} thin films}.
\newblock \bibinfo{journal}{Appl. Phys. Lett.} \bibinfo{volume}{81},
  \bibinfo{pages}{2587--2589}.
\newblock \DOIprefix\doi{10.1063/1.1508418}.
\bibitem[{{Perez~de~Lara} et~al.(2010){Perez~de~Lara}, Alija, Gonzalez, Velez,
  Martin and Vicent}]{LARA10}
\bibinfo{author}{{Perez~de~Lara}, D.}, \bibinfo{author}{Alija, A.},
  \bibinfo{author}{Gonzalez, E.M.}, \bibinfo{author}{Velez, M.},
  \bibinfo{author}{Martin, J.I.}, \bibinfo{author}{Vicent, J.L.},
  \bibinfo{year}{2010}.
\newblock \bibinfo{title}{Vortex ratchet reversal at fractional matching fields
  in kagom{\'{e}}like array with symmetric pinning centers}.
\newblock \bibinfo{journal}{Phys. Rev. B} \bibinfo{volume}{82},
  \bibinfo{pages}{174503}.
\newblock \DOIprefix\doi{10.1103/PhysRevB.82.174503}.
\bibitem[{Poccia et~al.(2015)Poccia, Baturina, Coneri, Molenaar, Wang,
  Bianconi, Brinkman, Hilgenkamp, Golubov and Vinokur}]{POCC15}
\bibinfo{author}{Poccia, N.}, \bibinfo{author}{Baturina, T.I.},
  \bibinfo{author}{Coneri, F.}, \bibinfo{author}{Molenaar, C.G.},
  \bibinfo{author}{Wang, X.R.}, \bibinfo{author}{Bianconi, G.},
  \bibinfo{author}{Brinkman, A.}, \bibinfo{author}{Hilgenkamp, H.},
  \bibinfo{author}{Golubov, A.A.}, \bibinfo{author}{Vinokur, V.M.},
  \bibinfo{year}{2015}.
\newblock \bibinfo{title}{Critical behavior at a dynamic vortex
  insulator-to-metal transition}.
\newblock \bibinfo{journal}{Science} \bibinfo{volume}{349},
  \bibinfo{pages}{1202--1205}.
\newblock \DOIprefix\doi{10.1126/science.1260507}.
\bibitem[{Reichhardt and Olson~Reichhardt(2007)}]{REIC07a}
\bibinfo{author}{Reichhardt, C.}, \bibinfo{author}{Olson~Reichhardt, C.J.},
  \bibinfo{year}{2007}.
\newblock \bibinfo{title}{Vortex molecular crystal and vortex plastic crystal
  states in honeycomb and kagom\'e pinning arrays}.
\newblock \bibinfo{journal}{Phys. Rev. B} \bibinfo{volume}{76},
  \bibinfo{pages}{064523}.
\newblock \DOIprefix\doi{10.1103/PhysRevB.76.064523}.
\bibitem[{Reichhardt et~al.(2015)Reichhardt, Ray and
  {Olson~Reichhardt}}]{REIC15}
\bibinfo{author}{Reichhardt, C.}, \bibinfo{author}{Ray, D.},
  \bibinfo{author}{{Olson~Reichhardt}, C.J.}, \bibinfo{year}{2015}.
\newblock \bibinfo{title}{Reversible ratchet effects for vortices in conformal
  pinning arrays}.
\newblock \bibinfo{journal}{Phys. Rev. B} \bibinfo{volume}{91},
  \bibinfo{pages}{184502}.
\newblock \DOIprefix\doi{10.1103/physrevb.91.184502}.
\bibitem[{{Savel'ev} and Nori(2002)}]{SAVE02}
\bibinfo{author}{{Savel'ev}, S.}, \bibinfo{author}{Nori, F.},
  \bibinfo{year}{2002}.
\newblock \bibinfo{title}{Experimentally realizable devices for controlling the
  motion of magnetic flux quanta in anisotropic superconductors}.
\newblock \bibinfo{journal}{Nat. Mater.} \bibinfo{volume}{1},
  \bibinfo{pages}{179--184}.
\newblock \DOIprefix\doi{10.1038/nmat746}.
\bibitem[{Sefrioui et~al.(2001)Sefrioui, Arias, Gonz\'alez, L\'eon, Santamaria
  and Vicent}]{SEFR01}
\bibinfo{author}{Sefrioui, Z.}, \bibinfo{author}{Arias, D.},
  \bibinfo{author}{Gonz\'alez, E.M.}, \bibinfo{author}{L\'eon, C.},
  \bibinfo{author}{Santamaria, J.}, \bibinfo{author}{Vicent, J.L.},
  \bibinfo{year}{2001}.
\newblock \bibinfo{title}{Vortex liquid entanglement in irradiated
  {YBa$_2$Cu$_3$O$_{7-\delta}$} thin films}.
\newblock \bibinfo{journal}{Phys. Rev. B} \bibinfo{volume}{63},
  \bibinfo{pages}{064503}.
\newblock \DOIprefix\doi{10.1103/PhysRevB.63.064503}.
\bibitem[{Silhanek et~al.(2006)Silhanek, Gillijns, Moshchalkov, Zhu, Moonens
  and Leunissen}]{SILH06}
\bibinfo{author}{Silhanek, A.V.}, \bibinfo{author}{Gillijns, W.},
  \bibinfo{author}{Moshchalkov, V.V.}, \bibinfo{author}{Zhu, B.Y.},
  \bibinfo{author}{Moonens, J.}, \bibinfo{author}{Leunissen, L.H.A.},
  \bibinfo{year}{2006}.
\newblock \bibinfo{title}{Enhanced pinning and proliferation of matching
  effects in a superconducting film with a {P}enrose array of magnetic dots}.
\newblock \bibinfo{journal}{Appl. Phys. Lett.} \bibinfo{volume}{89},
  \bibinfo{pages}{152507}.
\newblock \DOIprefix\doi{10.1063/1.2361172}.
\bibitem[{Silhanek et~al.(2010)Silhanek, Van~de Vondel and
  Moshchalkov}]{SILH10R}
\bibinfo{author}{Silhanek, A.V.}, \bibinfo{author}{Van~de Vondel, J.},
  \bibinfo{author}{Moshchalkov, V.V.}, \bibinfo{year}{2010}.
\newblock \bibinfo{title}{Guided vortex motion and vortex ratchets in
  nanostructured superconductors}, in: \bibinfo{editor}{Moshchalkov, V.V.},
  \bibinfo{editor}{Wördenweber, R.}, \bibinfo{editor}{Lang, W.} (Eds.),
  \bibinfo{booktitle}{Nanoscience and Engineering in Superconductivity}.
  \bibinfo{publisher}{Springer}, \bibinfo{address}{Heidelberg}, pp.
  \bibinfo{pages}{1--24}.
\newblock \DOIprefix\doi{10.1007/978-3-642-15137-8}.
\bibitem[{Sochnikov et~al.(2010)Sochnikov, Shaulov, Yeshurun, Logvenov and
  Bozovic}]{SOCH10}
\bibinfo{author}{Sochnikov, I.}, \bibinfo{author}{Shaulov, A.},
  \bibinfo{author}{Yeshurun, Y.}, \bibinfo{author}{Logvenov, G.},
  \bibinfo{author}{Bozovic, I.}, \bibinfo{year}{2010}.
\newblock \bibinfo{title}{Large oscillations of the magnetoresistance in
  nanopatterned high-temperature superconducting films}.
\newblock \bibinfo{journal}{Nat. Nanotechnol.} \bibinfo{volume}{5},
  \bibinfo{pages}{516--519}.
\newblock \DOIprefix\doi{10.1038/nnano.2010.111}.
\bibitem[{Swiecicki et~al.(2012)Swiecicki, Ulysse, Wolf, Bernard, Bergeal,
  Briatico, Faini, Lesueur and Villegas}]{SWIE12}
\bibinfo{author}{Swiecicki, I.}, \bibinfo{author}{Ulysse, C.},
  \bibinfo{author}{Wolf, T.}, \bibinfo{author}{Bernard, R.},
  \bibinfo{author}{Bergeal, N.}, \bibinfo{author}{Briatico, J.},
  \bibinfo{author}{Faini, G.}, \bibinfo{author}{Lesueur, J.},
  \bibinfo{author}{Villegas, J.E.}, \bibinfo{year}{2012}.
\newblock \bibinfo{title}{Strong field-matching effects in superconducting
  {YBa$_2$Cu$_3$O$_{7-\delta}$} films with vortex energy landscapes engineered
  via masked ion irradiation}.
\newblock \bibinfo{journal}{Phys. Rev. B} \bibinfo{volume}{85},
  \bibinfo{pages}{224502}.
\newblock \DOIprefix\doi{10.1103/physrevb.85.224502}.
\bibitem[{Tafuri(2024)}]{TAFU24R}
\bibinfo{author}{Tafuri, F.}, \bibinfo{year}{2024}.
\newblock \bibinfo{title}{Josephson junctions}, in:
  \bibinfo{editor}{Chakraborty, T.} (Ed.), \bibinfo{booktitle}{Encyclopedia of
  Condensed Matter Physics (Second Edition)}. \bibinfo{publisher}{Academic
  Press}, \bibinfo{address}{Oxford}, pp. \bibinfo{pages}{616--631}.
\newblock \DOIprefix\doi{10.1016/B978-0-323-90800-9.00145-1}.
\bibitem[{Tafuri and Kirtley(2005)}]{TAFU05R}
\bibinfo{author}{Tafuri, F.}, \bibinfo{author}{Kirtley, J.R.},
  \bibinfo{year}{2005}.
\newblock \bibinfo{title}{Weak links in high critical temperature
  superconductors}.
\newblock \bibinfo{journal}{Rep. Progr. Phys.} \bibinfo{volume}{68},
  \bibinfo{pages}{2573--2663}.
\newblock \DOIprefix\doi{10.1088/0034-4885/68/11/r03}.
\bibitem[{Tinchev(1996)}]{TINC96}
\bibinfo{author}{Tinchev, S.S.}, \bibinfo{year}{1996}.
\newblock \bibinfo{title}{Properties of {YBCO} weak links prepared by local
  oxygen-ion induced modification}.
\newblock \bibinfo{journal}{Physica C} \bibinfo{volume}{256},
  \bibinfo{pages}{191--198}.
\newblock \DOIprefix\doi{10.1016/0921-4534(95)00615-x}.
\bibitem[{Trastoy et~al.(2014)Trastoy, Malnou, Ulysse, Bernard, Bergeal, Faini,
  Lesueur, Briatico and Villegas}]{TRAS14}
\bibinfo{author}{Trastoy, J.}, \bibinfo{author}{Malnou, M.},
  \bibinfo{author}{Ulysse, C.}, \bibinfo{author}{Bernard, R.},
  \bibinfo{author}{Bergeal, N.}, \bibinfo{author}{Faini, G.},
  \bibinfo{author}{Lesueur, J.}, \bibinfo{author}{Briatico, J.},
  \bibinfo{author}{Villegas, J.E.}, \bibinfo{year}{2014}.
\newblock \bibinfo{title}{Freezing and thawing of artificial ice by thermal
  switching of geometric frustration in magnetic flux lattices}.
\newblock \bibinfo{journal}{Nat. Nanotechnol.} \bibinfo{volume}{9},
  \bibinfo{pages}{710--715}.
\newblock \DOIprefix\doi{10.1038/nnano.2014.158}.
\bibitem[{Villegas et~al.(2006)Villegas, Montero, Li and Schuller}]{VILL06}
\bibinfo{author}{Villegas, J.E.}, \bibinfo{author}{Montero, M.I.},
  \bibinfo{author}{Li, C.P.}, \bibinfo{author}{Schuller, I.K.},
  \bibinfo{year}{2006}.
\newblock \bibinfo{title}{Correlation length of quasiperiodic vortex lattices}.
\newblock \bibinfo{journal}{Phys. Rev. Lett.} \bibinfo{volume}{97},
  \bibinfo{pages}{027002}.
\newblock \DOIprefix\doi{10.1103/physrevlett.97.027002}.
\bibitem[{Villegas et~al.(2003)Villegas, Savel'ev, Nori, Gonzalez, Anguita,
  Garc\'{i}a and Vicent}]{VILL03}
\bibinfo{author}{Villegas, J.E.}, \bibinfo{author}{Savel'ev, S.},
  \bibinfo{author}{Nori, F.}, \bibinfo{author}{Gonzalez, E.M.},
  \bibinfo{author}{Anguita, J.V.}, \bibinfo{author}{Garc\'{i}a, R.},
  \bibinfo{author}{Vicent, J.L.}, \bibinfo{year}{2003}.
\newblock \bibinfo{title}{A superconducting reversible rectifier that controls
  the motion of magnetic flux quanta}.
\newblock \bibinfo{journal}{Science} \bibinfo{volume}{302},
  \bibinfo{pages}{1188--1191}.
\newblock \DOIprefix\doi{10.1126/science.1090390}.
\bibitem[{Vinckx et~al.(2007)Vinckx, Vanacken, Moshchalkov,
  M{\'a}t{\'e}fi-Tempfli, M{\'a}t{\'e}fi-Tempfli, Michotte, Piraux and
  Ye}]{VINC07}
\bibinfo{author}{Vinckx, W.}, \bibinfo{author}{Vanacken, J.},
  \bibinfo{author}{Moshchalkov, V.V.}, \bibinfo{author}{M{\'a}t{\'e}fi-Tempfli,
  S.}, \bibinfo{author}{M{\'a}t{\'e}fi-Tempfli, M.}, \bibinfo{author}{Michotte,
  S.}, \bibinfo{author}{Piraux, L.}, \bibinfo{author}{Ye, X.},
  \bibinfo{year}{2007}.
\newblock \bibinfo{title}{High field matching effects in superconducting {Nb}
  porous arrays catalyzed from anodic alumina templates}.
\newblock \bibinfo{journal}{Physica C} \bibinfo{volume}{459},
  \bibinfo{pages}{5--10}.
\newblock \DOIprefix\doi{10.1016/j.physc.2007.04.194}.
\bibitem[{Van~de Vondel et~al.(2005)Van~de Vondel, Silva, Zhu, Morelle and
  Moshchalkov}]{VAND05}
\bibinfo{author}{Van~de Vondel, J.}, \bibinfo{author}{Silva, C.C.D.},
  \bibinfo{author}{Zhu, B.Y.}, \bibinfo{author}{Morelle, M.},
  \bibinfo{author}{Moshchalkov, V.V.}, \bibinfo{year}{2005}.
\newblock \bibinfo{title}{Vortex-rectification effects in films with periodic
  asymmetric pinning}.
\newblock \bibinfo{journal}{Phys. Rev. Lett.} \bibinfo{volume}{94},
  \bibinfo{pages}{057003}.
\newblock \DOIprefix\doi{10.1103/PhysRevLett.94.057003}.
\bibitem[{Wambaugh et~al.(1999)Wambaugh, Reichhardt, Olson, Marchesoni and
  Nori}]{WAMB99}
\bibinfo{author}{Wambaugh, J.F.}, \bibinfo{author}{Reichhardt, C.},
  \bibinfo{author}{Olson, C.J.}, \bibinfo{author}{Marchesoni, F.},
  \bibinfo{author}{Nori, F.}, \bibinfo{year}{1999}.
\newblock \bibinfo{title}{Superconducting fluxon pumps and lenses}.
\newblock \bibinfo{journal}{Phys. Rev. Lett.} \bibinfo{volume}{83},
  \bibinfo{pages}{5106--5109}.
\newblock \DOIprefix\doi{10.1103/physrevlett.83.5106}.
\bibitem[{Ward et~al.(2006)Ward, Notte and Economou}]{WARD06}
\bibinfo{author}{Ward, B.W.}, \bibinfo{author}{Notte, J.A.},
  \bibinfo{author}{Economou, N.P.}, \bibinfo{year}{2006}.
\newblock \bibinfo{title}{Helium ion microscope: A new tool for nanoscale
  microscopy and metrology}.
\newblock \bibinfo{journal}{J. Vac. Sci. Techn. B} \bibinfo{volume}{24},
  \bibinfo{pages}{2871--2874}.
\newblock \DOIprefix\doi{10.1116/1.2357967}.
\bibitem[{Weber(2003)}]{WEBE03R}
\bibinfo{author}{Weber, H.W.}, \bibinfo{year}{2003}.
\newblock \bibinfo{title}{Irradiation}, in: \bibinfo{editor}{Cardwell, D.A.},
  \bibinfo{editor}{Ginley, D.S.} (Eds.), \bibinfo{booktitle}{Handbook of
  Superconducting Materials}. \bibinfo{publisher}{{IOP} Publishing},
  \bibinfo{address}{Bristol}, pp. \bibinfo{pages}{407--418}.
\newblock \DOIprefix\doi{10.1201/9781420034202}.
\bibitem[{Welp et~al.(2002)Welp, Xiao, Jiang, Vlasko-Vlasov, Bader, Crabtree,
  Liang, Chik and Xu}]{WELP02}
\bibinfo{author}{Welp, U.}, \bibinfo{author}{Xiao, Z.L.},
  \bibinfo{author}{Jiang, J.S.}, \bibinfo{author}{Vlasko-Vlasov, V.K.},
  \bibinfo{author}{Bader, S.D.}, \bibinfo{author}{Crabtree, G.W.},
  \bibinfo{author}{Liang, J.}, \bibinfo{author}{Chik, H.}, \bibinfo{author}{Xu,
  J.M.}, \bibinfo{year}{2002}.
\newblock \bibinfo{title}{Superconducting transition and vortex pinning in {Nb}
  films patterned with nanoscale hole arrays}.
\newblock \bibinfo{journal}{Phys. Rev. B} \bibinfo{volume}{66},
  \bibinfo{pages}{212507}.
\newblock \DOIprefix\doi{10.1103/physrevb.66.212507}.
\bibitem[{W\"ordenweber et~al.(2004)W\"ordenweber, Dymashevski and
  Misko}]{WORD04}
\bibinfo{author}{W\"ordenweber, R.}, \bibinfo{author}{Dymashevski, P.},
  \bibinfo{author}{Misko, V.R.}, \bibinfo{year}{2004}.
\newblock \bibinfo{title}{Guidance of vortices and the vortex ratchet effect in
  high-{$T_c$} superconducting thin films obtained by arrangement of antidots}.
\newblock \bibinfo{journal}{Phys. Rev. B} \bibinfo{volume}{69},
  \bibinfo{pages}{184504}.
\newblock \DOIprefix\doi{10.1103/physrevb.69.184504}.
\bibitem[{Xue et~al.(2018)Xue, Ge, He, Zharinov, Moshchalkov, Zhou, Silhanek
  and Van~de Vondel}]{XUE18}
\bibinfo{author}{Xue, C.}, \bibinfo{author}{Ge, J.Y.}, \bibinfo{author}{He,
  A.}, \bibinfo{author}{Zharinov, V.S.}, \bibinfo{author}{Moshchalkov, V.V.},
  \bibinfo{author}{Zhou, Y.H.}, \bibinfo{author}{Silhanek, A.V.},
  \bibinfo{author}{Van~de Vondel, J.}, \bibinfo{year}{2018}.
\newblock \bibinfo{title}{Tunable artificial vortex ice in nanostructured
  superconductors with a frustrated kagome lattice of paired antidots}.
\newblock \bibinfo{journal}{Phys. Rev. B} \bibinfo{volume}{97},
  \bibinfo{pages}{134506}.
\newblock \DOIprefix\doi{10.1103/PhysRevB.97.134506}.
\bibitem[{Yun et~al.(2000)Yun, Pedarnig, R\"ossler, B\"auerle and
  Obradors}]{YUN00}
\bibinfo{author}{Yun, S.H.}, \bibinfo{author}{Pedarnig, J.D.},
  \bibinfo{author}{R\"ossler, R.}, \bibinfo{author}{B\"auerle, D.},
  \bibinfo{author}{Obradors, X.}, \bibinfo{year}{2000}.
\newblock \bibinfo{title}{In-plane and out-of-plane resistivities of vicinal
  {Hg-1212} thin films}.
\newblock \bibinfo{journal}{Appl. Phys. Lett.} \bibinfo{volume}{77},
  \bibinfo{pages}{1369--1371}.
\newblock \DOIprefix\doi{10.1063/1.1289489}.

\end{thebibliography}

\end{document}